\documentclass[12pt]{article}
\usepackage[margin=2 cm]{geometry}
\usepackage{comment}
\usepackage{pdfpages}
\usepackage{graphicx}
\usepackage[font={small,it}]{caption}
\usepackage{mwe}
\usepackage[nottoc]{tocbibind}
\usepackage{amsmath,amssymb,extarrows,mathtools,graphicx,subfigure,setspace}
\usepackage{cite}
\usepackage{stackengine}
\usepackage{tikz}\usetikzlibrary{calc}
\usepackage[compat=1.0.0]{tikz-feynman}
\usepackage{braket}
\usepackage{epsfig}
\usepackage[section]{placeins}
\usepackage{slashed}
\usepackage{color}
\usepackage{caption}
\usepackage{amsmath}
\usepackage{hyperref}
\makeatother

\newcommand{\be}{\begin{equation}}
\newcommand{\bea}{\begin{eqnarray}}
\newcommand{\eea}{\end{eqnarray}}
\newcommand{\ba}{\begin{array}}
\newcommand{\ea}{\end{array}}
\newcommand{\ee}{\end{equation}}
\newcommand{\bes}{\begin{equation*}}
\newcommand{\beas}{\begin{eqnarray*}}
\newcommand{\eeas}{\end{eqnarray*}}
\newcommand{\bas}{\begin{array*}}
\newcommand{\eas}{\end{array*}}
\newcommand{\ees}{\end{equation*}}

\numberwithin{equation}{section}

\begin{document}

\onehalfspacing
\vfill
\begin{titlepage}
\vspace{10mm}

\begin{center}

\vspace*{10mm}
\vspace*{1mm}
{\Large  \textbf{String scattering amplitudes and mutual information in confining backgrounds: the partonic behavior}} 
 \vspace*{1cm}
 
{$\text{Mahdis Ghodrati}^{a, b}$},

\vspace*{8mm}
{ \textsl{
$^a $ International Centre for Theoretical Physics Asia-Pacific,
University of Chinese Academy of Sciences, 100190 Beijing, China}
} 

{ \textsl{
$^b$ Asia Pacific Center for Theoretical Physics,
Pohang University of Science and Technology, Pohang 37673, Republic of Korea}
}

 \vspace*{0.4cm}
\textsl{e-mails: {\href{mahdisghodrati@ucas.ac.cn}{mahdisghodrati@ucas.ac.cn}}}
 \vspace*{2mm}

\vspace*{1.7cm}

\end{center}

\begin{abstract}
Using AdS/CFT methods in various holographic confining backgrounds, we present the connections between the behaviors of string scattering amplitudes and mutual information. We specifically lay down the analogies between the logarithmic branch cut behavior of the string scattering amplitude in $4d$, $ \mathcal{A}_4 $,  at low and medium Mandelstam variable $s$, which is due to the dependence of the string tension on the holographic radial coordinate, and the branch cut behavior observed in mutual information and critical distance $D_c$ at low-cut-off variable $u_{KK}$. In both cases, as $s$ or $u_{KK}$ increases, the peaks of the branch cuts fade away in the form of $\text{Re}\lbrack \mathcal{A}_4 \rbrack \propto s^{-1}$. Next, the modular flow and modular Hamiltonian as the intermediary quantities will be used to further shed lights on this connection. We then explain how mutual information itself can act as a probe of chaos in various scenarios. For two examples of Compton scattering (of a photon entangled with another photon off a target) and also the decay of a highly excited string into two tachyons, the pattern of entanglement entropy and the change in the mutual information would be examined. Finally, the kink-kink and kink-antikink scatterings as simple models of scattering in confining geometries are emploted to probe the fractal structures in the scatterings of topological defects. 
 \end{abstract}

\end{titlepage}

\tableofcontents

\section{Introduction}

In 1968, the SLAC laboratory observed that at high energies and fixed angles, the hadronic scattering amplitudes fall off based on a power low in the Mandelstam variable $s$ (the center of mass energy), and not, as previously imagined, in an exponential form of $s$. That observation was incompatible with the expectation from string theory, therefore, further studies were needed from the theory perspective.  On the other hand, using holography, interesting connections between string scattering amplitude and entanglement entropy have been found which could help in solving this puzzle, for instance check \cite{Seki:2014pca, Hubeny:2014zna, Semenoff:2011ng,Hubeny:2014kma}.  Additionally, the correspondences between strings and black holes discussed for instance in \cite{Chen:2021dsw} pointed out that a lot of information about quantum gravity and quantum information problems from the structures of string amplitude could be obtained. In fact, as there are links between the chaotic behavior of the excited strings and the chaoticity of back holes, one could expect that one of the best physical quantity to employ first would the scattering amplitude, since the perturbative string theory has been formulated using it.

Building on these connections \cite{Seki:2014pca,Hubeny:2014zna,Semenoff:2011ng,Hubeny:2014kma,Polchinski:2002jw,Andreev:2004sy,Hatta:2007he,Pire:2008zf,BallonBayona:2007qr,Brower:2000rp,Brunner:2015oqa, Mahapatra:2019uql, Jain:2020rbb}, we can now move to the richer mixed correlation measures, such as negativity and mutual information and further establish the correlations between the behaviors of  scattering amplitudes and the pattern of quantum entanglement in holography.

In \cite{Rosenhaus:2021xhm}, the scattering amplitude of highly excited strings has been studied within the DDF (Del Giudice, Di Vecchia, Fubini \cite{DelGiudice:1971yjh}) formalism whereby the excited string is produced by photons that are being scattered off of an initial tachyon repeatedly. We use this formalism to analyze the connections between the behaviors of string scatterings and the mixed quantum information measures.

\begin{figure}[ht!]
 \centering
  \includegraphics[width=5.75cm] {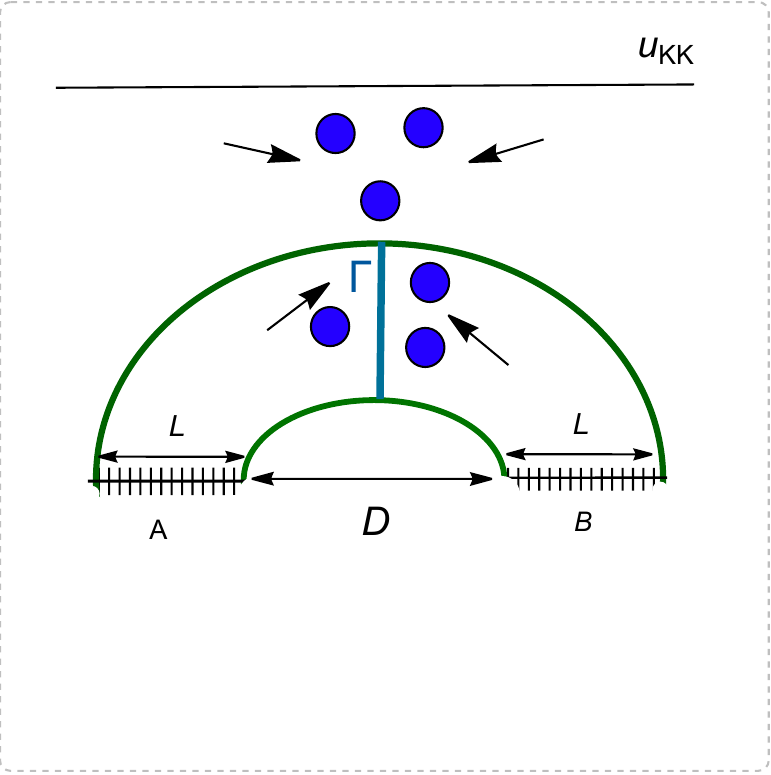} 
    \includegraphics[width=5.75cm] {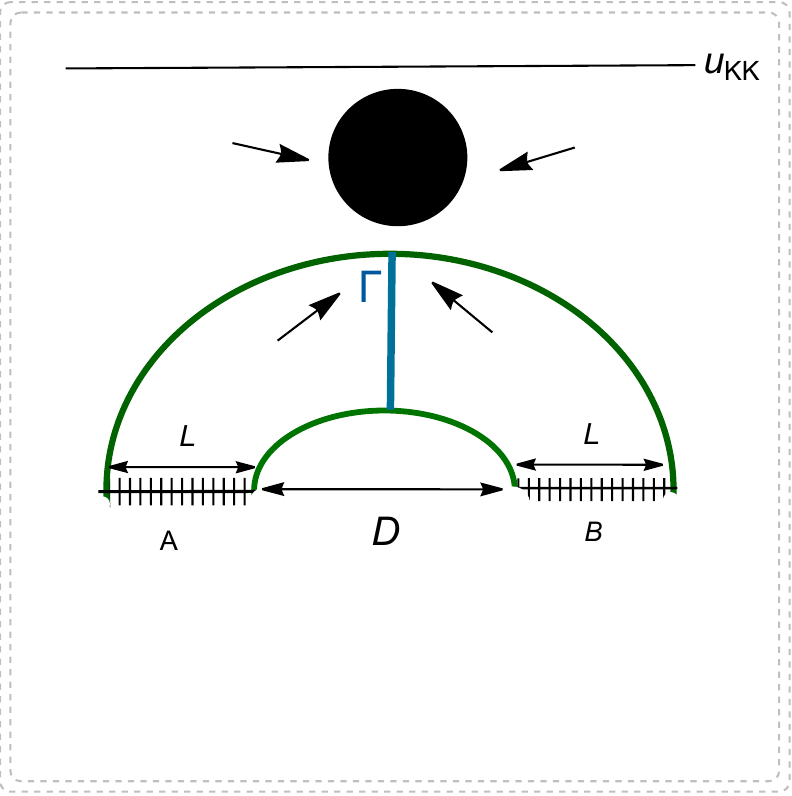} 
        \includegraphics[width=5.75cm] {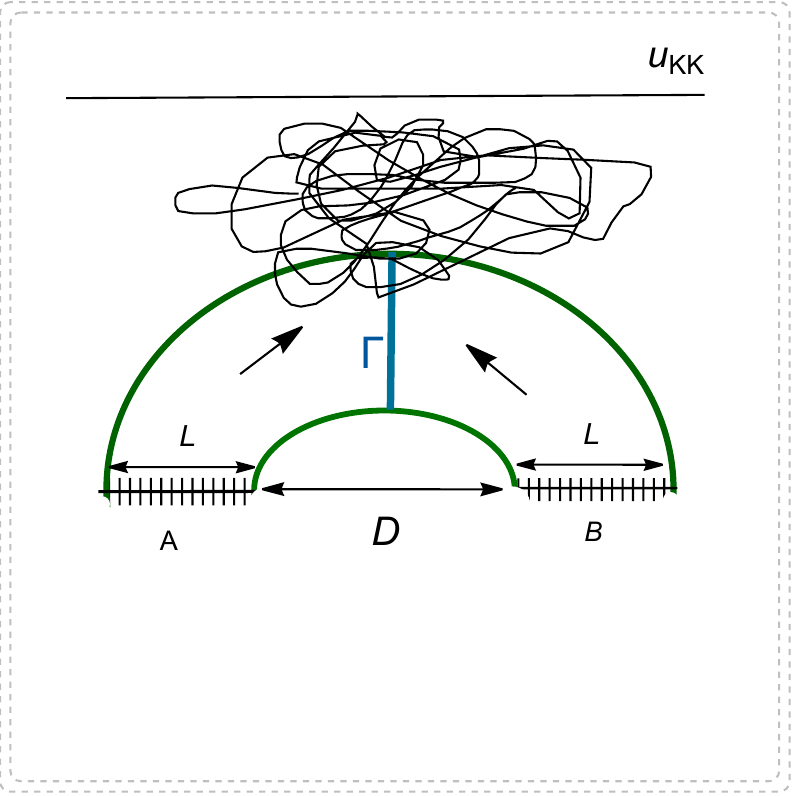} 
  \caption{In the left part, the scattering of ``classical" pinballs, in the middle, the ``quantum" and highly chaotic scattering process off of a black hole, and in the right part, the scattering off of highly excited strings in a confining background with an end-wall positioned at $u_{KK}$ are shown.}
 \label{fig:wedgeScatter}
\end{figure}

In fig. \ref{fig:wedgeScatter}, three examples of scattering problems in the confining background of a mixed entanglement wedge constructed from two infinite strips with width $L$ and distance $D$ is shown.
This figure is similar to figure 1 of \cite{Gross:2021gsj}, but here we specifically consider a confining background with an end wall positioned at $u_{KK}$ constructed from the mixed correlations of two strips. The classical chaos can be studied by the model of the scattering of pinballs (left part of fig. \ref{fig:wedgeScatter}), where the angle of the outgoing particle is very sensitive to the impact parameter. This angle is related to several other parameters of the setup as well. However, this scattering process is not sensitive to the ``quantum" mutual information of the two systems $A$ and $B$. 

In the middle part of fig. \ref{fig:wedgeScatter}, the particles scattering off of a black hole system in the background of a confining geometry is sketched. In this scenario, any perturbation in any parameters of the system could cause a large change in the state of the outgoing Hawking quanta, where the out-of-time-order correlator (OTOC) grows exponentially. Therefore, this setup can depict the very chaotic nature of black holes. The level of chaos in this case is higher than the pinball example in the sense that it has both the ``classical" and ``quantum" chaoticity. In this setup, depending on the size of the black hole, the position of the wall $u_{KK}$, and the lengths of $L$ and $D$, various phases could exist.

In the right part of fig. \ref{fig:wedgeScatter}, the sketch of a complicated and highly excited open string structure which is positioned in the confining and mixed background is shown. As will be shown later, in this case there would be an exponentially large number of internal states and dependence on the parameters of the system. So, the scattering amplitude displays high level of chaos, and high sensitivity on various parameters of the system. Also, based on the size of the parameters of this setup relative to each other and the strength of the correlations among the two systems, at any specific configuration, a particular saddle would be dominant. So, a very rich phase structure for this system could be envisioned.

The entanglement entropy has already been employed in various works to find connections between energy conditions and quantum information theory. For example in \cite{Balakrishnan:2017bjg}, using the second order shape deformation of the geometric entanglement entropy, a conjectural lower bound on the expectation value of the components of energy momentum tensor has been found which has been dubbed the Quantum Null Energy Condition (QNEC). In addition, the properties of the modular Hamiltonian under shape deformations for any state $\psi$ can help in this regard \cite{Faulkner:2016mzt,Balakrishnan:2017bjg}.

The difference between our setup and figure. 1 of \cite{Gross:2021gsj} is that here the background is constructed from the entanglement of “two mixed correlated subsystems”, and there is also a wall at the end of the geometry. The modular flow in holography defines the proper internal time flow for an observer in this system. Our observation here is related to the result of \cite{Bousso:2020yxi}, where the Connes cocycle (CC) flow, kink transform and the bulk Weyl tensor shock have all been connected with each other.

For this system, one can imagine the non-equilibrium modular flows and the “internal modular time”  (instead of the real physical time) \cite{deBoer:2022zps, Arias:2020qpg},  which describes the pattern of entanglement. The modular flows \cite{DeBoer:2019kdj} from these two sources of mixed subsystems hit the wall at the end of the geometry, and then slide or reflect back \cite{Apolo:2020qjm}, constructing a chaotic structure with those observed peaks, where by increasing the Mandelstam variables $s$ these peaks would decay. 

The measure of mutual information in quantum field theories can detect the patterns of entanglement in a more precise way as it is UV regulated. Therefore, the mutual information is a superior measure to extract the universalities in the structures of the entanglement of the system.  It could also depict the intrinsic behaviors of the strings in a more exhaustive way. In addition, as pointed out in \cite{Chen:2022cmr}, compared to entanglement entropy, mutual information is less constrained by the symmetries and hence it can illustrate better the details, such as the full spectrum of the CFTs directly related to the behaviors of the fundamental strings.

Specifically, the singularities in the scattering amplitudes, which have been observed in \cite{Bianchi:2021sug}, would form logarithmic branch cuts (rather than poles). These erratic behaviors and branch cuts, which are more pronounced for small $s$ and fade away for larger $s$, can be explained by the creation of bound states in a confining background. The same erratic behavior and peak-structures due to these bound states can be detected through mutual information as well, as we have observed in our numerical studies in \cite{Ghodrati:2021ozc, Ghodrati:2022kuk}.

In \cite{Bianchi:2021sug}, specifically, the partonic behavior of scattering processes at larger and medium ranges of Mandelstam variable $s$ in the holographic confining backgrounds have been explored. This observation was based on the works of Polchinski and Strassler \cite{Polchinski:2001tt}, whose their main result was writing the $4d$ amplitude in terms of the $10d$ scattering amplitude as
\begin{gather}\label{eq:A4rel1}
\mathcal{A}_4 (s,t,u) \propto \int dr \sqrt{-g} \times \mathcal{A}_{10} \lbrack \tilde{s}(r), \tilde{t}(r), \tilde{u}(r) \rbrack,
\end{gather}
where $\mathcal{A}_{10}$ is the $10d$ string amplitude, $r$ is the holographic radial coordinate, and $\Psi (r)$ is the wave function of the scattered state. In the holographic QCD models with a non-trivial dilaton field such as Maldacena-Nuñez \cite{Maldacena:2000mw} or Witten-QCD \cite{Witten:1997ep,Seiberg:1994aj}, where the dilaton field depends on the radial coordinate, the above relation should instead be written as
\begin{gather}\label{eq:A4rel2}
\mathcal{A}_4 (s,t,u) = \int_{u_{KK} }^\infty  du \sqrt{-g} e^{-2 \phi}  \Big( \prod_{i=1}^{4}  \psi_i(u)  \Big)  \mathcal{A}_{10} \lbrack \tilde{s}(r), \tilde{t}(r), \tilde{u}(r) \rbrack.
\end{gather}

Unlike theories such as $\mathcal{N}=4$ SYM, the spectrum of QCD is discrete, which is the result of the IR end-wall and the boundary condition that it imposes.  This discrete spectrum can also affect the behaviors of quantum information correlation measures.  Moreover, the entanglement entropy and mutual information can also detect the asymptotic freedom where the gauge couplings and the binding energies become smaller at that point.  On the other hand, as found in \cite{Bianchi:2021sug}, the scattering amplitude itself could act as a measure which catches the passage from soft to hard scattering, which is the ``bending" trajectory in the confining models.  The ``quantized" behaviors of such binding energies could also be detected at lower energies.

In this setup, most of the closed strings reside near the end-wall, where instead of being vanished, this accumulation causes the wave function to get peaked there which might seem counterintuitive. This feature shows its signatures on the entanglement patterns and mixed correlations saddles as well. Based on such partonic behavior of hadron scattering  \cite{Bianchi:2021sug, Kang:2004jd, Pire:2008zf}, one would expect that the entanglement patterns could detect such behavior as well and the direct connections between these two quantities could be constructed more firmly, as we demonstrate here.

This paper is organized as follows. In section \ref{sec:sec2}, we discuss the fractal structures coming from the phase transitions and chaos in confining geometries due to the presence of the wall. These structures have been captured by the mutual information and critical distance $D_c$, noted in our previous works \cite{Ghodrati:2015rta, Ghodrati:2022kuk, Ghodrati:2022hbb,Ghodrati:2021ozc}. In section \ref{sec:sec3}, similar to the results of \cite{Bianchi:2021sug}, we explore scattering amplitude in $10d$ and $4d$, the structures of its zeros and poles and  their statistics which prove the appearance of chaos. We also discuss how the structures of the zeros would behave the same for both mutual information and string scattering amplitude.

In section \ref{sec:sec4}, we use the modular flow as instrumental concepts to further establish the relations between the mutual information and string scattering amplitude. In section \ref{sec:sec5}, we will show that the mutual information itself can detect the chaotic behaviors.  There, we also discuss quantum scattering on a leaky torus and using this geometrical structure, we further explore the chaotic behaviors of strings and also the chaotic spread of mutual information in confining geometries. In section \ref{sec:sec6}, we analyze two examples of Compton scattering with a witness and also the decay of a highly excited string into two tachyons, where for each case we discuss the behavior of entanglement entropy or mutual information during these processes. Next, in section \ref{sec:sec7}, as another example of chaotic structure in confining geometries, we consider the kink-kink and kink-antikink scattering, which similar to the behavior of $D_c$ and mutual information in confining geometries, they show fractal structures with periodic patterns. Finally, we give concluding remarks in section \ref{sec:sec8}.

\section{Critical distance in confining backgrounds}\label{sec:sec2}

\begin{figure}[ht!]
 \centering
  \includegraphics[width=8cm] {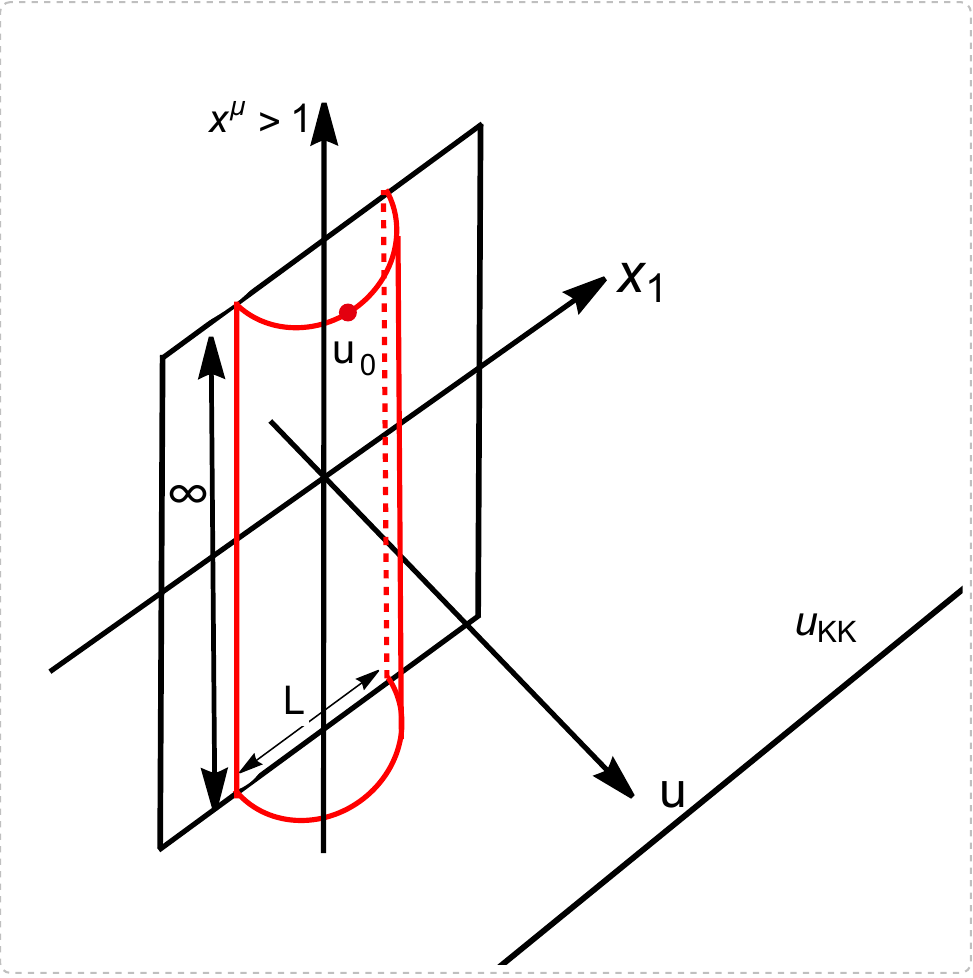} 
  \caption{The setup of one strip and its corresponding Ryu-Takayanagi surface in a confining background.}
 \label{fig:stripStringFrame}
\end{figure}

In \cite{Kol:2014nqa,Brandhuber:1998er}, it has been shown that, in order to calculate the entanglement entropy for a general confining background with the geometry of
\begin{gather}
ds^2= \alpha(u) \lbrack \beta(u) du^2 + dx^\mu dx_\mu \rbrack + g_{ij} d \theta^i d\theta^j, \nonumber\\
 u_{KK} < u < \infty, \ \ x^\mu ( \mu =0,1, . . . ,d), \ \  \theta^i ( i=d+2, . . . 9),
\end{gather}
in terms of the minimum of the holographic radial coordinate, the length of a strip and its corresponding Ryu-Takayanagi surface, $u_0$, can be written as
\begin{gather}
L(u_0)= 2 \int_{u_0}^\infty du \sqrt{\frac{\beta(u) }{ \frac{\mathtt{H}(u) }{\mathtt{H}(u_0) } -1} }, \nonumber\\
S(u_0) = \frac{V_0}{ 2 G_N} \int_{u_0}^\infty du \sqrt{ \frac{\beta(u) \mathtt{H}(u) }{1- \frac{\mathtt{H}(u_0) }{\mathtt{H}(u)} } }, 
\end{gather}
where here $\mathtt{H}(u)= e^{-4 \phi} V^2_{\text{int}} \alpha(u)^d$ and $V_{\text{int} } = \int d \vec{\theta} \sqrt{\text{det} \lbrack g_{ij}\rbrack}$.
The corresponding setup is shown in figure \ref{fig:stripStringFrame}.

For two strips of width $L$ and the distance $D$, the mutual information would be
\begin{gather}
I(D,L)= 2 S(L)-S(D) - S(2L+D),
\end{gather}
and the critical distance $D_c$ is where $I(D_c,L)=0$.  The parameter $D_c$ (as can be noted in figure \ref{fig:wedgeScatter}) has been used in our previous works \cite{Ghodrati:2022hbb, Ghodrati:2022kuk,Ghodrati:2020vzm, Ghodrati:2021ozc,Ghodrati:2019hnn} to probe the quantum correlation properties of different geometries.

\begin{figure}[ht!]
 \centering
  \includegraphics[width=12cm] {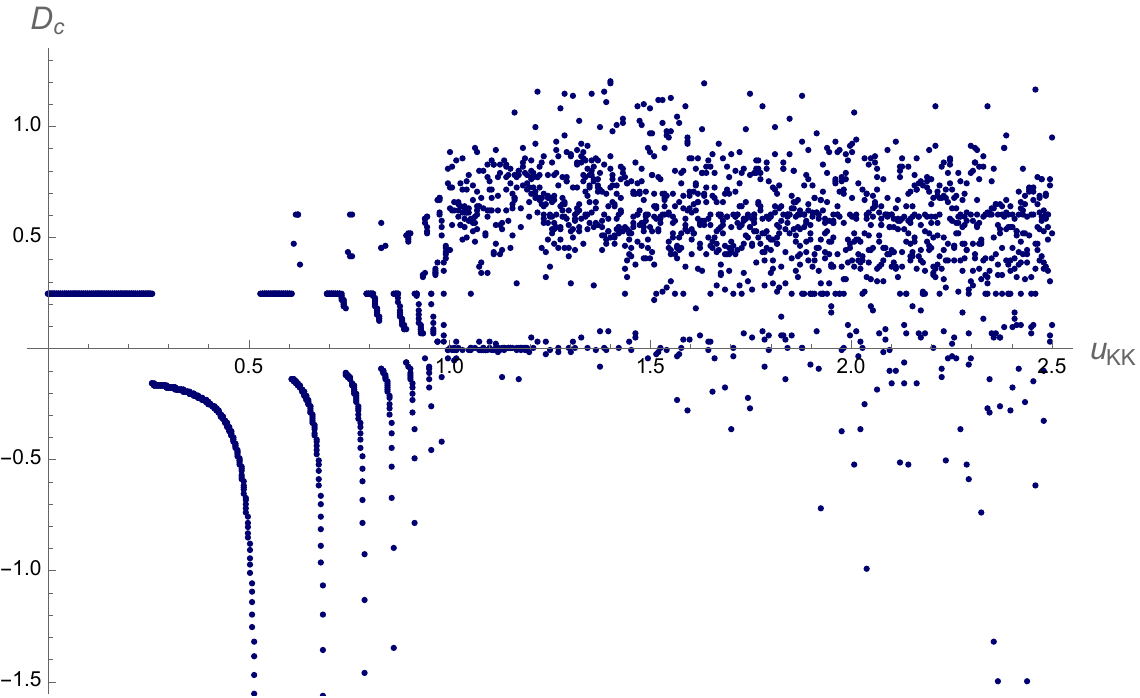}  
  \caption{The behavior of the critical distance, $D_c$, which comes from the mutual information between two mixed strips, versus the position of the IR wall, $u_{KK}$, in the Witten-QCD background. This figure can be compared with figure 6 of \cite{Bianchi:2021sug} on the basis of the ``real parts of the amplitude".  The behavior of the branch cuts and the decay at large $s$ are very similar to the pattern of MI and $D_c$ in this model.  }
 \label{fig:wedgewittenQCD}
\end{figure}

\begin{figure}[ht!]
 \centering
  \includegraphics[width=9cm] {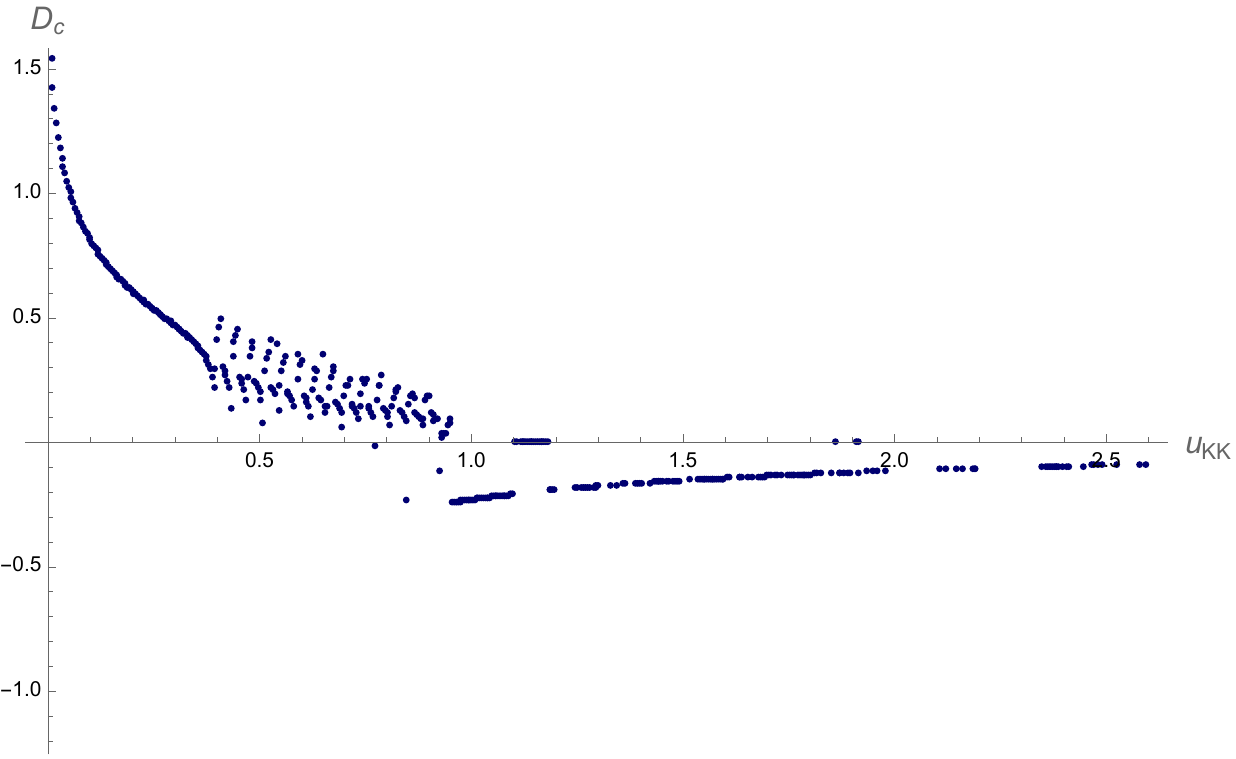} 
  \caption{Left: The relationship between the critical distance $D_c$ among two strips, versus $u_{KK}$, in the background of Klebanov-Tseytlin geometry.}
 \label{fig:wedgeKT}
\end{figure}

In \cite{Ghodrati:2021ozc}, we found that, in the background of confining geometries, the critical distance between two strips as a measure of mixed correlations shows singularities at smaller values of $u_{KK}$ and statistically smooths out in bigger $u_{KK}$. This behavior is especially pronounced in the Witten-QCD geometry, i.e., figure 19 of \cite{Ghodrati:2021ozc}, which is shown again here in figure \ref{fig:wedgewittenQCD}. The same jumps can also be seen in the background of Klebanov-Tseytlin geometry \cite{Klebanov:2000nc, Klebanov:2000hb}, as presented here in figure \ref{fig:wedgeKT}.

The potential in the Witten-QCD geometry can be written as $V(r) \sim r^{4/3}$ follows a warped metric structure and confinement at large distances, would grow softer than a linear potential. The geometry is cigar-like and the time direction is periodic, avoiding the conical singularity.

Similar to the results of \cite{Bianchi:2021sug}, it could be seen that, in the Witten-QCD background, the peaks of $D_c$ which correspond to the peaks of mutual information and mixed correlation measures are bigger in the medium (or rather smaller) regions of $u_{KK}$ which correspond to the smaller values of Mandelstam variable $s$. For the bigger values of $s$ (and bigger $t$ where the angle is fixed), which correspond to bigger $u_{KK}$, the peaks fade away. This is the result of asymptotic freedom where the gauge couplings and the binding energies of the bound states are small. However, from figures \ref{fig:wedgewittenQCD} and \ref{fig:wedgeKT}, it could be seen that there is no peak at larger $u_{KK}$, which is due to the fact that the binding energy, length and the mass of the strings become smaller in those setups. Note also similar to \cite{Mahapatra:2019uql}, and the case of tripartite information $I^{ \lbrack n=3 \rbrack}$, the measures could be negative with abrupt, singular drops.

Note that this is similar to the result of \cite{Mukhopadhyay:2021wmu} which described the kink-antikink scattering in a quantum vacuum. In their model, they have two fields, $\varphi(t,x)$ with mass $m$ and a scalar field $\psi(t,x)$ with mass $\mu$ and coupling constant $\lambda$. The exchange of the energy between the classical field configuration and quantum bath is similar to the procedure of strings hitting the end-wall in confining models and therefore similar patterns in the observables are expected.
When the coupling constant $\lambda$ increases, from figures 4 and 5 of \cite{Mukhopadhyay:2021wmu}, which are the plots of $E_\phi$ versus $t$, one can see that the cascading decay becomes an oscillatory decay and the bound states become long-lived, weakly radiating objects. This is similar to our figure \ref{fig:wedgewittenQCD}, where the bound states and breather-like phase decay to radiating phase. Also, the phase shifts in kink-antikink scattering is related to the memory of past interactions being restored in the quantum mutual information. Also, both quantities show nonlinear behavior. Plus, as there is a critical distance and a threshold for mutual information, there would also be a certain energy threshold where only above it, the kinks and antikinks can interact significantly. We will explain the kink-antikink model further in section \ref{sec:sec7}.

 \begin{figure}[ht!]
 \centering
  \includegraphics[width=11cm] {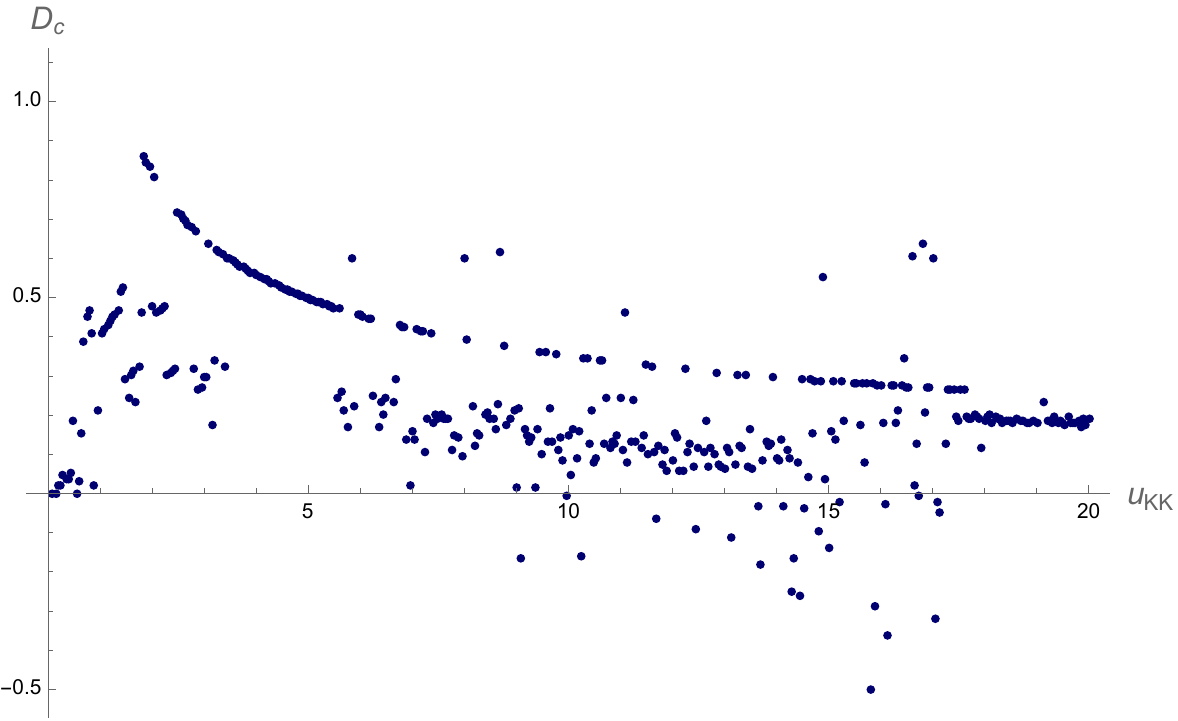} 
  \caption{Top: The relationship between the critical distance $D_c$  versus $u_{KK}$, in the background of Sakai-Sugimoto geometry. Again, one can see that the peaks become less pronounced as $u_{KK}$ increases and fade away at large values of $s$. The decay behavior at large $u_{KK}$ is a power law.}
 \label{fig:SakaiDc}
\end{figure}

The behavior of the critical distance $D_c$ versus $u_{KK}$, and the potentail in the background of Sakai-Sugimoto \cite{Sakai:2004cn} is shown in figure \ref{fig:SakaiDc}. As this figure has been created from the ``real" part of the entanglement entropy, $Re(S(u_0))$ versus the real part of the size of the strips $Re(L(u_0))$, one would expect that, similarly in the ``real" part of the amplitude, at bigger values of $s$ or $u$, a decaying power low behavior should be observed. This is indeed what one can see from the behaviors in figure \ref{fig:SakaiDc}. The relation for the amplitude follows $\text{Re}\lbrack \mathcal{A}_4 \rbrack \sim s^{-1}$ and $\text{Im} \lbrack \mathcal{A}_4 \rbrack \sim s^{2- \Delta/2}$, where $\Delta$ is the sum of the scaling dimensions of the operators that contribute as $\prod_{i=1} \psi_i(r) \sim r^{- \Delta} $, ($\Delta$ can be replaced by the twist operator in QCD models).

 Also note that among the models we consider here, Sakai-Sugimoto is the one with most similarities with the model studied in \cite{Hahne:2023dic} for kink-antikink scattering. As found there that there would be two separate regions with highest amount of radiation which are also the most chaotic regions, in the mutual information or $D_c$, we also notice two separate peaks one around $u_{KK}\simeq 2$ and one around $u_{KK} \simeq 17$ which correspond to $v \simeq 0.39$ and $v \simeq 0.075$ in the case of kink-antikink scattering. Note that bigger $u_{KK}$ corresponds to lower velocities.

\begin{figure}[ht!]
 \centering
  \includegraphics[width=9cm] {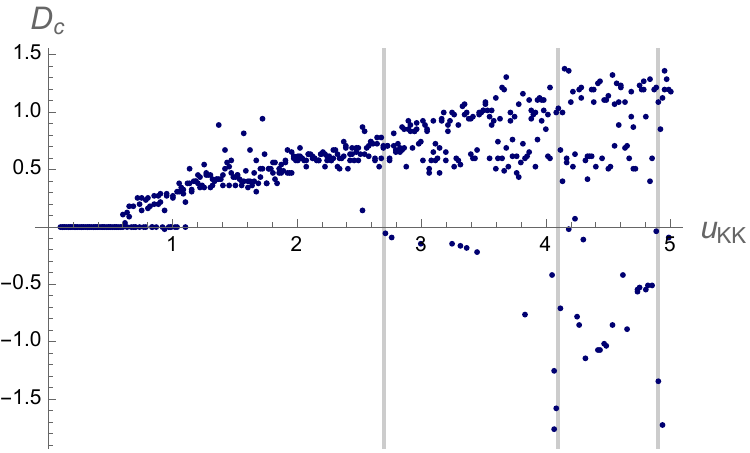} 
  \caption{The relationship between the critical distance $D_c$ among the two strips versus $u_{KK}$, in the background of AdS-soliton geometry, and its potential. One can compare these diagrams with the plots of Crofton form presented in \cite{Ghodrati:2021ozc,Ghodrati:2022kuk} as they can highlight the specific characteristics of each geometry. }
 \label{fig:AdSSoliton}
\end{figure}

As shown in figure \ref{fig:AdSSoliton}, in the background of AdS-soliton, the transition from regular to chaotic regimes can be clearly noted in the behavior of $D_c$. This plot should be compared with figure 7 of \cite{Bianchi:2023uby} where the transition from regular to random  happens at large $u_{KK}$s. Note that bigger values of $u_{KK}$ correspond to smaller values of the discrete levels or ``eigenvalues". The peaks become less pronounced as $u_{KK}$ increases and fade away at large values of $s$. This is similar to the case of  \cite{Bianchi:2023uby} which as energy and $s$ increase, the singularities fade away.  However, the general behaviors of mutual information or $D_c$ versus $u_{KK}$ are not exactly similar in different geometries and depends on other factors of the backgrounds such as flux, genus, etc.

For the case of hard-wall confining geometries such as Witten-QCD, the decay at large $u_{KK}$ is a decreasing power law similar to its behavior in terms of $s$.  As explained in \cite{Bianchi:2023uby}, the cause of this behavior in the large energy domains is the asymptotic freedom, which when being approached, the corresponding gauge coupling becomes weaker and the binding energy of the bound states vanishes. This is similar to the case where as $u_{KK}$ increases and the asymptotic freedom is being approached, the binding energy decreases, which based on the first law of entanglement entropy leads to smaller peaks in the entanglement entropy, mutual information and also $D_c$.

For the case of AdS-soliton however, as one can analyze the geometry using the potential and also its Crofton form as in \cite{Ghodrati:2021ozc,Ghodrati:2022kuk}, the well in low holographic coordinates $u$ is positive. For bigger $u$s, the Crofton form while being negative approaches zero which is different from the case of hard-wall models such as Witten-QCD and Sakai-Sugimoto model. Therefore, using the Crofton form, one can track the behavior of critical distance $D_c$. For the case of AdS-Soliton, the bound states could be created at larger $u_{KK}$s while for smaller $u_{KK}$s, the critical distance $D_c$ would be zero, as the Crofton form $\omega$ is infinite there. 

For interpreting this result, one can visualize the spreads of wave functions between the two entangled strips in spacetime, where they interfere with each other, collide with the wall at the end of the confining geometry, ripple back and create the chaotic structures and the peaks which fade away at bigger values of $s$.  These waves are being sourced from the two entangled mixed subregions creating the ``modular chaos" which form the shape of the peaks in the plots of mutual information.

Also, it has been demonstrated, for instance in \cite{DeBoer:2019kdj}, that the modular chaos structures similar to entanglement (and mutual information) can reconstruct the holographic bulk geometry.  Further, in \cite{Bianchi:2021sug}, the connections between ``chaos" and the peaks structures of ``string amplitude" have been discussed. Consequently, one would think that the structures of the peaks of the string amplitude and the structures of entanglement should have connections with each other through ``modular chaos”, as we will prove.

The disruption of entanglement entropy, mutual information, or $D_c$ in this case could be a probe of chaos. From figure \ref{fig:AdSSoliton} however, it could be seen that the ``edge of chaos" is not very clear. This can also be related to the result of \cite{carroll2019mutualinformationedgechaos}, where they showed that the performance of a reservoir computer at the edge of chaos, where the largest Lyapunov exponent in a dynamical system shifts from negative to positive, might be optimum or poor,  as the actual performance would depend on various other parameters as well. It was previously believed that the greater mutual information, or critical distance, between the fit signal $h(t)$ and the training signal $g(t)$ in reservoir computers leads to an improved performance. Also, at the edge of chaos, the modular flow and the string scattering amplitudes could fluctuate.

\section{String scattering amplitude in confining backgrounds}\label{sec:sec3}

Now, turning to the string amplitude, as shown in \cite{Bianchi:2021sug}, the $10d$ amplitude can be found as
\begin{gather}
\mathcal{A}_{10} = 4^4 \sum_{n=0}^\infty (-1)^n \frac{(n+1)^2}{(n!)^2} \frac{(1+c_+^4 + c_-^4)}{\alpha' s/4 - (n+1)} \times \nonumber\\
\frac{\Gamma \left( (n+1) c_+\right)  \Gamma \left( (n+1) c_- \right) }{\Gamma \left (1-(n+1) c_+\right) \Gamma \left(1-(n+1)c_-\right)  } \equiv \sum_{n=0}^\infty \frac{\mathcal{R} (\theta)}{\alpha' s/4-(n+1)},
\end{gather}
where $c_{\pm}= \frac{1}{2} (1 \pm \cos \theta) $. Here, bigger values of $\alpha'$ correspond to bigger values of $u_{KK}$ and higher energy levels. In terms of $s$ and the angle $\theta$, the amplitude can be written as
\begin{gather}
\mathcal{A}_{10}=\frac{1}{32} {\alpha' }^4 s^4 (\cos (2 \theta )+7)^2 \frac{\Gamma \left(-\frac{1}{4} ( {\alpha'} s )\right)}{\Gamma \left(\frac{ {\alpha'} s }{4}+1\right)}\times \nonumber\\
\frac{\Gamma \left(-\frac{1}{8}  {\alpha'} s  (\cos (\theta )-1)\right)}{\Gamma \left(\frac{1}{8} (- {\alpha'} s + \text{$ {\alpha'} s $} \cos (\theta )+8)\right)} \times \frac{\Gamma \left(\frac{1}{8} {\alpha'} s  (\cos (\theta )+1)\right)}{\Gamma \left(1-\frac{1}{8}  {\alpha'} s  (\cos (\theta )+1)\right)}.
\end{gather}

\begin{figure}[ht!]
 \centering
  \includegraphics[width=8.6cm] {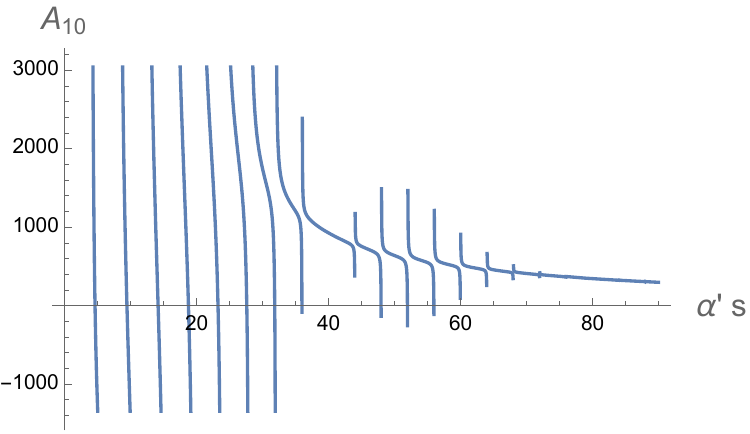} 
    \includegraphics[width=8.6cm] {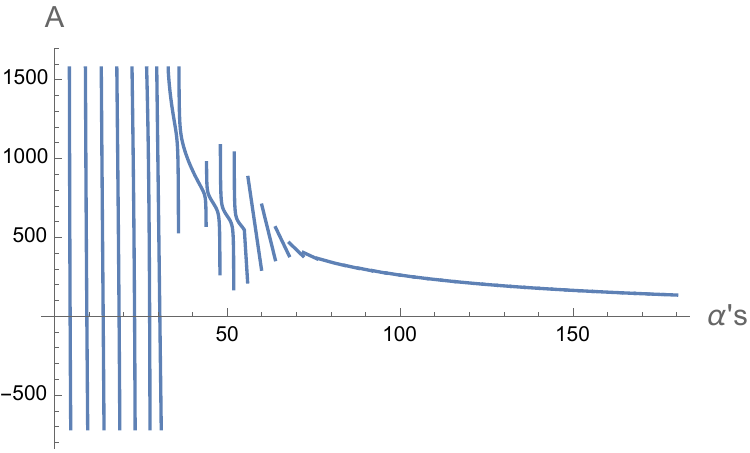} 
  \caption{The relationship between $A_{10}$ versus $\alpha' s$ for $\cos \theta =0.8$. The weakening pattern of the branch cut singularities at large Mandelstam variable $s$ is similar to the behavior of $D_c$ versus $u_{KK}$ in confining backgrounds.}
 \label{fig:Aalpha22}
\end{figure}

\begin{figure}[ht!]
 \centering
  \includegraphics[width=12.5cm] {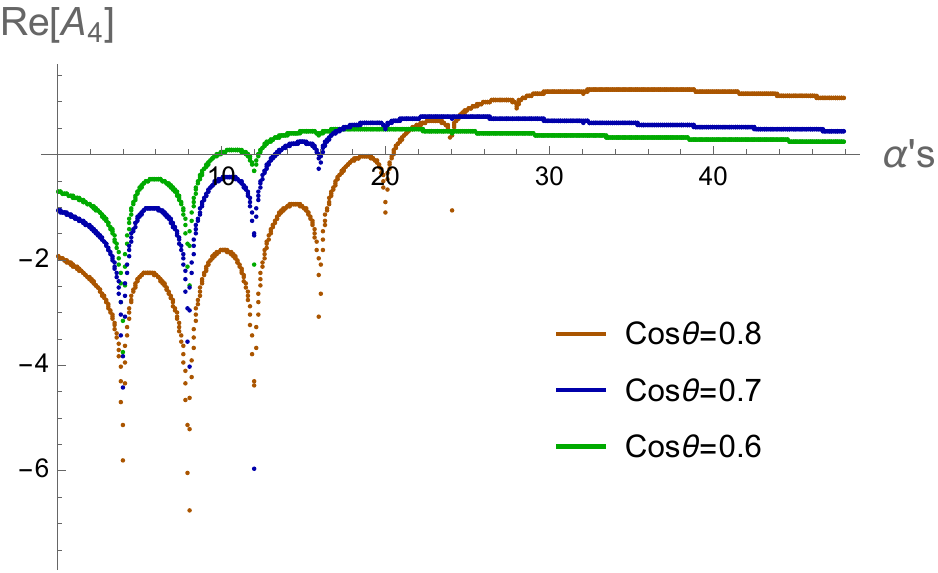} 
  \caption{The relationship between the real part of $\lbrack \mathcal{A}_4 \rbrack$ versus $\alpha' s$ for different values of $\theta$, in the hard wall confining model, created based on the results of \cite{Bianchi:2021sug}. The main behaviors are similar to the case of Witten-QCD. The logarithmic branch cut singularities at lower $s$, and the damping patterns at larger $s$, i.e,. $\lbrack \mathcal{A}_4 \rbrack \sim s^{-1}$, are similar to the behaviors of the critical distance $D_c$ (coming from MI), observed in various confining backgrounds.}
 \label{fig:SakaiDc2}
\end{figure}

 In \cite{Bianchi:2021sug}, the behavior of $\mathcal{A}_{10}$ versus the Mandelstam variable $s$ in confining holographic backgrounds such as Witten-QCD has been shown. The specific confining backgrounds considered there included hard-wall, soft-wall, and Witten-QCD models, showing similar singularity patterns to entanglement entropy and mutual information, as shown in figure \ref{fig:Aalpha22}.
  Then, using the relations \ref{eq:A4rel1} and \ref{eq:A4rel2}, $\mathcal{A}_{10}$ can be integrated and the scattering amplitude in $4d$ can be derived. As for our main figure, the behavior of $\text{Re} \lbrack \mathcal{A}_4 \rbrack$ for different values of $\theta$ is shown in figure \ref{fig:SakaiDc2}, which should be compared with the behavior of $D_c$ in various confining backgrounds.

The reason for such peculiar behaviors in generating these singularities in the form of the branch cuts at low energies is the dependence of the string tension on the holographic radial coordinate.  It can be noted that in $10d$ or $11d$ geometries, the amplitudes drop faster compared to lower $d$ geometries, which is similar to the flattening of singularities in the mixed-correlation measures in higher dimensions.

\begin{figure}[ht!]
 \centering
  \includegraphics[width=6.9cm] {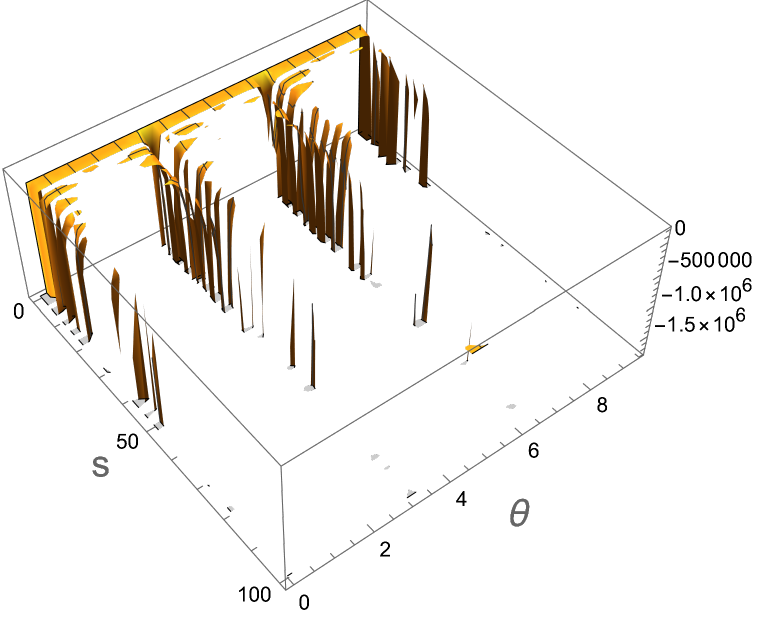}  \ \ \ \ \ \ \ 
    \includegraphics[width=6.9cm] {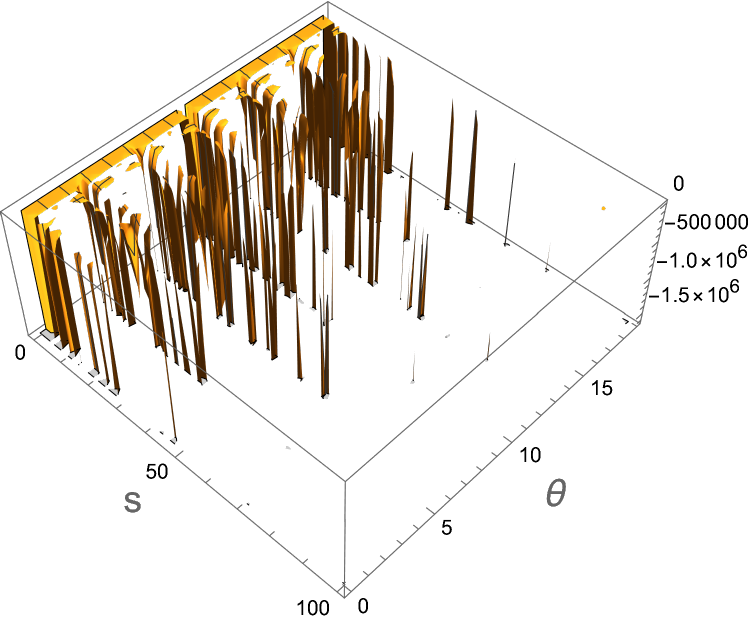} 
  \caption{The $3d$ plot of the $10d$ scattering amplitude, $A_{10}$, versus $s$ and $\theta$. We set $\alpha'=1$ here. }
 \label{fig:stheta}
\end{figure}

The connections between $s$ and $\theta$ have been further explored in figure \ref{fig:stheta} as in the interval of $0 <\theta < 3\pi$, nearly four branch cuts, and in the interval of $0 <\theta < 6\pi$, nearly seven branch cuts can be found. Owing to the fact that the geometry is $10d$, we are allowed to consider such values for the angle $\theta$ as we are considering scattering from highly excited strings in $10d$. Since we can consider bigger spins in the structures of chaotic and excited strings and specifically as we consider higher dimensions, it would be necessary  for $\theta$ to take bigger values than $2\pi$. In general, with the upper bound of $n \pi$, $n+1$ branch cuts would be present.  Also, it can be seen that for bigger $s$, the branch cuts become dispersed, which is also the case for the mixed-correlation quantum information measures such as mutual information and $D_c$.

\begin{figure}[ht!]
 \centering
  \includegraphics[width=8cm] {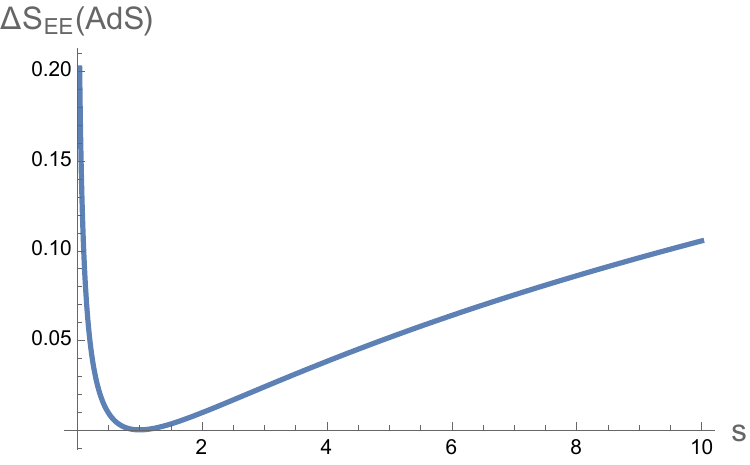} 
  \caption{The relationship between $\Delta S_{EE}$ and $s$  in pure AdS.}
 \label{fig:pureEE}
\end{figure}

One could see that in the simple case of fixed $t$, as shown in figure \ref{fig:pureEE}, $\Delta S_E \propto (\log s)^2 $. For the mixed correlation measures and in the presence of a hard wall which creates the ``four" main saddles \cite{Dong:2021oad,Ghodrati:2022hbb}, the behaviors become more interesting, which is shown in figures \ref{fig:wedgewittenQCD} and \ref{fig:wedgeKT}.   Since the parameter $D_c$ is related to the sum of several entanglement entropies, therefore, figure \ref{fig:pureEE} can be comparable to the case of the confining and mixed setup of figures \ref{fig:wedgewittenQCD} and \ref{fig:wedgeKT}. 

Note that the scattering process will always be dominated by the interior regions which is closer to the IR wall, and therefore the position of the wall is the main factor in determining the saddles. In fact, the mixed/random phase would be dominant closer to that wall which is one of the reason that the phase transitions can be probed effectively using MI and $D_c$ versus $u_{KK}$.

\begin{figure}[ht!]
 \centering
  \includegraphics[width=8cm] {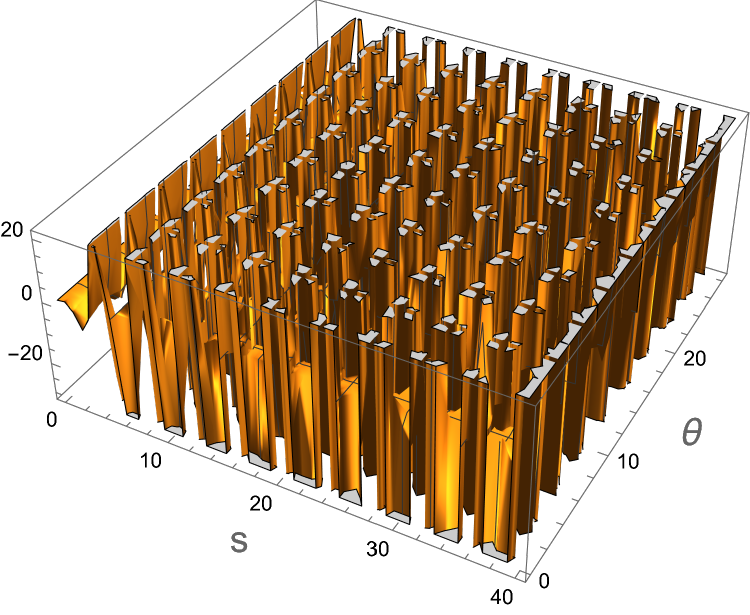} 
  \caption{The relationship between $A_{10}$ versus $s$ and $\theta$ for $\alpha=1$ in the fixed angle approximated limit.}
 \label{fig:fixedtheta}
\end{figure}

Now let's consider two interesting limits. The fixed angle limit approximation of $s \to \infty$ while $s/t$ is being fixed has been studied in \cite{Bianchi:2021sug}, where the behavior of the real part of the $A_{10}$ is shown in figure \ref{fig:fixedtheta}. Note that, this limit is not physical as there are imaginary contributions. Therefore, the behavior in this limit is not similar to the behavior of quantum mixed correlation measures unlike the case we saw for the full physical case of $A_{10}$ without this approximation.

\begin{figure}[ht!]
 \centering
  \includegraphics[width=7.2cm] {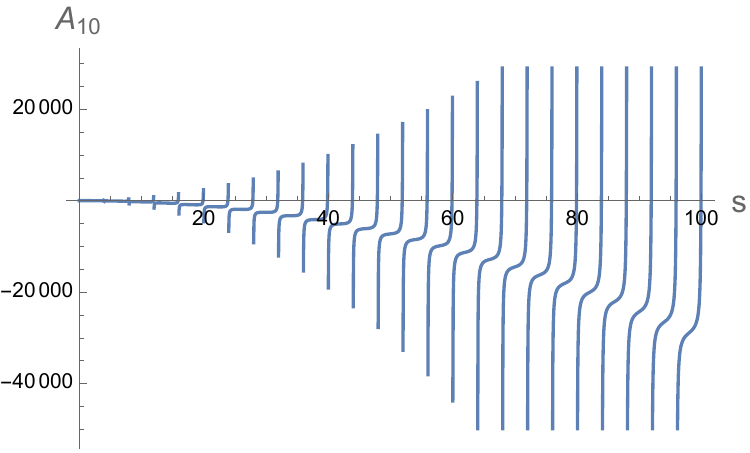} 
    \includegraphics[width=7.2cm] {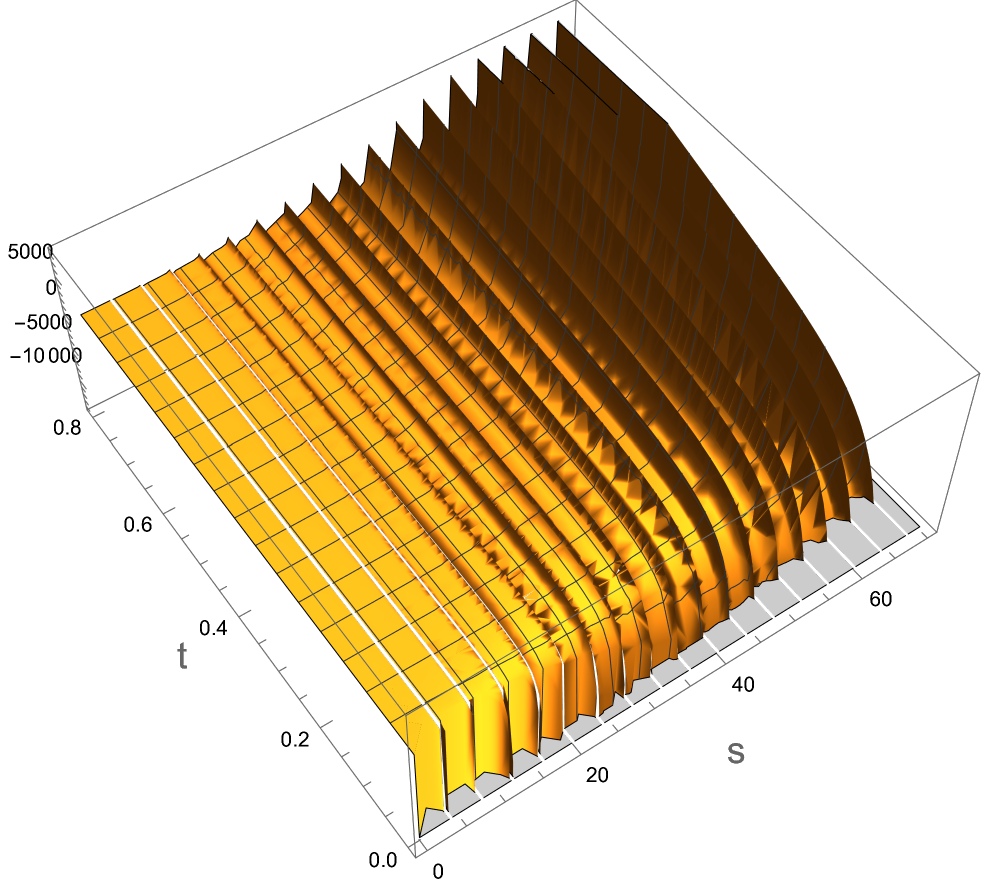} 
  \caption{The relationship between $A_{10}$ versus $s$ and $\theta$ for $\alpha=1$ in the Regge limit.}
 \label{fig:ReggeLimit}
\end{figure}

On the other hand, the Regge limit, where $s \to \infty$ while $t$ is fixed would give the figures of \ref{fig:ReggeLimit}. In this case the full phase diagram cannot be exposed while for fining the connections, having the full spectrum is needed. 

Other than comparing the structures of singularities and branch cuts in the plot of $D_c$ (MI) and string scattering amplitude as we have done previously, one could find more direct and analytical connections between various quantum information measures and scattering amplitudes.

In an older work, \cite{Seki:2014pca}, it has been shown that any interaction between particles would change the entanglement entropy (EE), and this change of EE has been directly linked to the scattering amplitude. This is also the case for all other mixed correlation measures such as mutual information and negativity. So the scattering amplitude should also able to track the ``four saddles" of the mixed correlations studied in \cite{Dong:2021oad,Ghodrati:2022hbb}.

Then, by employing the perturbation theory, the disc scattering amplitude of two heavy quark-anti-quark pairs which are being accelerated in the opposite directions has been found in \cite{Hubeny:2014zna}. So by taking the entanglement entropy of the quark and anti-quark which is approximately $S_{EE}= \sqrt{\lambda}$, where $\lambda$ is the 't Hooft coupling, the connections between these quantities can be demonstrated in a more rigorous and analytical way. In addition, the entanglement entropy can be written in terms of the Wilson loop, $\langle W \rangle$ as $S_E= (1- c \lambda \partial_\lambda ) \log \langle W \rangle$, (please check eq 1.1 \cite{Lewkowycz:2013laa} which is correct in general $d$-dimension,  and also  \cite{Seki:2014pca, Kitaev:2005dm}). Thus, at large 'Hooft couplings $\lambda$, the change of EE at leading order in $\lambda$ could be related to the gluon scattering amplitude as  \cite{Seki:2014pca}
 \begin{gather}
 \Delta S_E \sim \frac{(1-\frac{1}{2} c ) \sqrt{\lambda} }{8\pi} \left ( \log \frac{s}{t} \right )^2.
 \end{gather}

 Also, in \cite{Hubeny:2014zna}, the amplitude and then entanglement entropy between two accelerating quark and anti-quarks have been found from the 4-point function
\begin{gather}\label{eq:fourpoint}
g(x_f, \tilde{x}_f; x_i, \tilde{x}_i)= \int_0^\infty \frac{dT}{T} \int \lbrack dx^\mu \rbrack e^{i S \lbrack x, T \rbrack} \int_0^\infty \frac{d \tilde{T} }{\tilde{T} } \int \lbrack d\tilde{x}^\mu \rbrack e^{i \tilde{S} \lbrack \tilde{x} , \tilde{T} \rbrack }  W\lbrack x, \tilde{x} \rbrack,
\end{gather}
 where 
 \begin{gather}
 S=\int_0^1 d\sigma \left \lbrack \frac{(\dot{x}^\mu (\sigma))^2}{4T} - M^2 T + \frac{E}{2} ( x^1 \dot{x}^0 - x^0 \dot{x}^1 ) \right \rbrack, \nonumber\\
\tilde{S}=\int_0^1 d\sigma \left \lbrack \frac{(\dot{\tilde{x}}^\mu (\sigma))^2}{4\tilde{T}} - M^2 \tilde{T} - \frac{E}{2}(\tilde{x}^1 \dot{\tilde{x}}^0 - \tilde{x}^0 \dot{\tilde{x}}^1 ) \right \rbrack.
 \end{gather}
Also, $W\lbrack x, \tilde{x} \rbrack = \bra{0} \text{Tr} \ U\lbrack x \rbrack \tilde{U} \lbrack \tilde{x} \rbrack \ket{0}$, where
\begin{gather}
U \lbrack x \rbrack = \mathcal{T} e^{i \int_{- \infty}^\infty d \tau \lbrack A_\mu (x(\tau)) \dot{x}^\mu (\tau) + \sqrt{- \dot{x} (\tau)^2 } \phi^1(x(\tau)) \rbrack }, \nonumber\\
\tilde{U} \lbrack \tilde{x} \rbrack =  \left ( \mathcal{T} e^{i \int_{- \infty}^\infty d \tau \lbrack A_\mu (\tilde{x}(\tau)) \dot{\tilde{x}}^\mu (\tau) - \sqrt{- \dot{\tilde{x}} (\tau)^2 } \phi^1(\tilde{x}(\tau)) \rbrack } \right )^\dagger.
\end{gather} 
 
 So, one can see that increasing the mass $M$ would decrease $S$ and $\tilde{S}$, the four point function and the amplitude.

Similar analytical connections between the mixed correlation measures such as mutual information, negativity, entanglement and complexity of purification (EoP and CoP) on one hand, and the string scattering amplitude on the other hand are needed to be found. For instance the scattering fermions, the spin-momentum and helicity-momentum and also the variations of mutual information have been found in \cite{Fan:2017hcd} which could be extended to other measures as well. Also, the connections between string scattering and evolution of  Ryu-Takayanagi Surface and also reflected entropy have been studied in \cite{Jiang:2024noe}.

Another interesting observation is the links between the change of entanglement entropy and the Bremsstralung of radiative corrections, $\lambda \partial_\lambda \log \langle W \rangle $ in the case of accelerating quark-antiquarks pair, which is proportional to the Bremsstralung function \cite{Correa:2012at,Seki:2014pca}. So the ``spikes" that we have noticed during the phase transitions in the plot of $D_c$ could be related to K-lines in Bremsstralung radiation.

Note that our observation for the connections is more precise for the closed strings (glueballs). In the case of open strings, which describe mesons or baryons, there would be two kinds of entanglement entropy, which make finding the connections between mutual information and scattering amplitude more challenging. In that case, the entanglement entropy includes shares from the correlations between the strings' endpoints of the gluons (or flavor branes) and the correlations between the gluons themselves. This makes the framework and building the connections that we are interested somehow more complicated.

\section{Mutual information, modular Hamiltonian and binding energy}\label{sec:sec4}

In various papers, the entanglement entropy and mutual information have already been used to extract the universalities in the properties of renormalization group flows at various energies \cite{Myers:2010tj, Casini:2015woa}. Therefore, it is expected that they could also detect the quantized behaviors in low energies which occur due to the creation of binding energy, as found using the scattering amplitude in \cite{Bianchi:2021sug} and we inquired into in section \ref{sec:sec3}.  We could also use the modular Hamiltonian, modular flows and shape dependence, similar to the setup of \cite{Chen:2022cmr}, to formulate such connections. We should model the scattering process and its effects on the modular flows as the shape perturbations, similar to the setup of \cite{Chen:2022cmr}. Additionally, we can show that modular flows could also detect the partonic behaviors of \cite{Bianchi:2021sug} at low energies.

The mutual information between two spherical regions with radius $R_A$ and $R_B$ separated by a large distance $L$ could be written as \cite{Agon:2015ftl, Cardy:2013nua, Chen:2017hbk,Chen:2022cmr}
\begin{gather}
I_{A,B}= \mathcal{N}_\Delta \frac{\sqrt{\pi} \Gamma (2 \Delta+1) }{4 \Gamma (2 \Delta+\frac{3}{2} )} \left( \frac{R_A R_B }{L^2} \right )^{2\Delta} + . . . 
\end{gather}

In the above relation, $\Delta$ is the conformal dimension of an internal scalar primary operator that carries the mixed correlation between the spherical regions $A$ and $B$. The string scattering process would affect this scalar field and the weights of the ``carriers" and therefore the total correlations.

\begin{figure}[ht!]
 \centering
   \includegraphics[width=8.5cm] {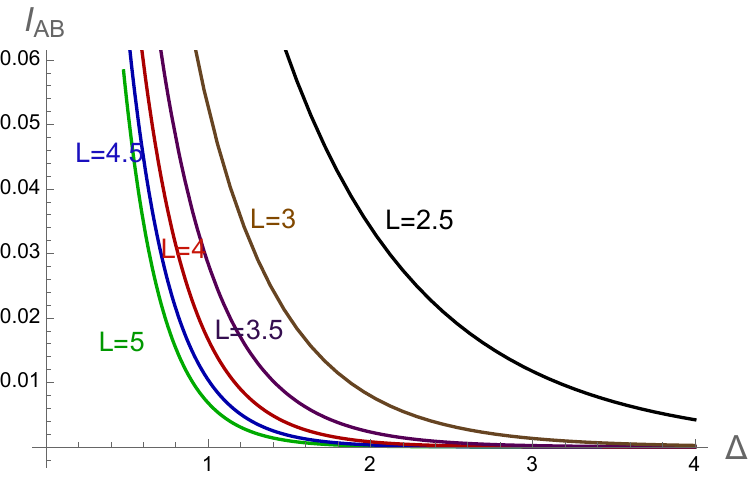} 
  \caption{The relationship between mutual information $I_{AB}$ versus the conformal dimension of an internal scalar primary operator $\Delta$, for various distance $L$ between the spherical regions. The same behavior is expected for the string scattering amplitude versus the mass of the field.}
 \label{fig:ILAB}
\end{figure}

The behavior of the mutual information $I_{AB}$ versus $\Delta$, for the case of $L>R_{A,B}$ has been shown in figure \ref{fig:ILAB}. One could see that by increasing $\Delta$, the mutual information decreases. Also, for any specific value of $\Delta$, the smaller distance $L$ corresponds to bigger mutual information $I_{AB}$ between the two spheres.

 The scaling dimension and the mass of the field $\phi$ are connected with each other as $\Delta=\frac{d}{2} + \sqrt{\frac{d^2}{4} + m^2 R^2}$. So, for the case of $L>R_{A,B}$, bigger $\Delta$ leads to bigger mass $m$ which makes the strings scattering process between the strings more difficult, leading to smaller amplitudes. This is consistent with the relation \ref{eq:fourpoint} as well. Furthermore, this can be checked from figure 3 of \cite{Bianchi:2019ywd}, where the interactions of string coherent states in the DDF formalism have been studied and the open string profiles at different average masses and their time evolutions were shown. There, one could see that by increasing the mass, their profile shrink, making the string scattering amplitude to decrease.  This can signal how scattering amplitude and mutual information are related.

Note that in other models of CFT such as $O(N)$, when the scaling dimension increases the system becomes more stable against the fluctuations of the magnetic field or the electric field. The charge in these models would  increase the anomalous dimension $\Delta$ \cite{Antipin:2021rsh}, and also breaks the correlations based on a decreasing power law function as in the form of figure \ref{fig:ILAB}.

The scattering process of an accelerating string or quark moving toward a sphere or  another sitting quark can be written as the perturbative model of shape deformation and the modular flow \cite{Chen:2022cmr}.  For any Hamiltonian $H$, the thermal density matrix can be written as $\rho_\beta = \frac{e^{- \beta H} }{\text{tr} (e^{- \beta H} )}$ and from the other side, for any invertible density matrix $\rho$, a thermal Hamiltonian can be constructed as $K_\rho \equiv - \text{log} \ \rho $, where $K_\rho$ is the modular Hamiltonian of the state $\rho$. So, any density matrix can be written as an exponential form of $\rho_V = c e^{- \mathcal{H}}$. The interesting thing about modular theory of these operators or \textit{Tomita-Takesaki theory} \cite{Takesaki:1970aki, Goheer:2003tx, Haag:1992hx} is that it can be constructed even for the case where the density matrices cannot be defined for the quantum systems.  The modular Hamiltonian can be written as the sum of a local part and a non-local part as $H = H_{\text{loc} } + H_{non-loc} $. They can in fact encode the complete entanglement data of the system.

For any Hilbert space $\mathcal{H}$ with a von Neumann algebra $\mathcal{A}$ and for a state $\ket{\Omega} \in \mathcal{H}$, the anti-linear operator $S_\Omega$ can be defined as \cite{Connes_1994}
\begin{gather}
S_\Omega ( a \ket{\Omega} ) = a^\dagger \ket{\Omega}, \ \ \ \ \ a \in \mathcal{A}.
\end{gather}

The modular operator $\Delta_\Omega \equiv S_\Omega^\dagger S_\Omega $ can be written using the operator $S_\Omega$ and then, using $\Delta_\Omega$, the modular Hamiltonian can be redefined as $K_\Omega \equiv - \log \Delta_\Omega $. Next, the \textit{modular flow} of operators can be set out via a map as
\begin{gather}
a  \mapsto e^{i K_\Omega t} a e^{- i K_{\Omega} t} = \Delta_\Omega^{- i t} a \Delta_\Omega^{it}.
\end{gather}

This is in fact a unitary operator which implements an ``internal time flow", which was the basis for us to connect the mutual information and string scattering amplitude. For the case of a system with more than one interval, the modular evolution is non-local, which causes the  ``teleportation" between the disjoint intervals \cite{Casini:2009vk}.  This can demonstrate that for connecting scattering amplitude and quantum information, one should in fact take the ``mixed correlation measures" for multi-regions rather than the entanglement entropy.

An interesting connection can be seen from the relation of modular Hamiltonian $H$ which can be written using the decomposition in terms of the eigenvectors as 
\begin{gather}
H(x,y) = 2\pi \sum_k \int_{- \infty}^\infty ds \ s \ \Psi_s^k(y) \Psi_s^{k*}(x),
\end{gather}
where $\Psi(s) \equiv \Psi (x^\mu (s))$, and $s_1$ and $s_2$ are length parameters along the spherical set $V$.  Also $\Psi (x)$s are the Dirac fields. So the sum of the scattered eigenvectors from the scattering process could reconstruct the modular Hamiltonian and the trajectories of modular flow.

Noe that the local part of Hamiltonian can be written as
\begin{gather}
H_{\text{loc}} = \pi i \left ( 2\left(\frac{dz(x)}{dx} \right)^{-1} \partial_x + \frac{d}{dx} \left ( \frac{dz(x)}{dx} \right )^{-1} \right ) \delta (x-y),
\end{gather}
and the non-local part can be written as
\begin{gather} \label{eq:poles}
H_{\text{non-loc} } = - 2\pi i \sum_{l, x_l(z(x)) \ne x} \frac{1}{(x-y)} \left( \frac{dz}{dy} \right )^{-1} \delta( y - x_l ( z(x) )),
\end{gather}
which mixes the finite number of points, i.e., $x_l (z)$, from each interval, which are related to the peaks of $D_c$. So, the singularities and the zeros in the plots of scattering amplitude and also $D_c$ are mostly related  to the ``non-local" correlations.

Due to the scattering process, the boundary of one of the subsystems changes as $\tilde{x}^\mu = x^\mu + \zeta^\mu$.  The strings that are being created from the vacuum for an instance in time would move along the geodesics which are determined by the linear response of the metric deformation.  Then, the scattering process, similar to \cite{Chen:2022cmr}, would be proportional to the deformation and ``modular flow", $\nabla^\mu \xi^\nu$, leading to the metric deformation $\delta g^{\mu \nu} =2 \nabla^\mu \xi^\nu $, which then leads to the change of the reduced density matrix as 
\begin{gather}
\delta \rho_A= U^\dagger \circ \rho_{\tilde{A}} \circ U - \rho_A\simeq \frac{1}{2} \int \delta g^{\mu \nu} \rho_A \tilde{T}_{\mu \nu}. 
\end{gather}

From the first law of entanglement entropy, one can deduce that, in general, the linear response of the entanglement entropy is proportional to the correlation of the contour integral of stress tensor with modular Hamiltonian as \cite{Chen:2022cmr} $\delta S_A \propto \oint_{z \sim \partial A} dz \langle T(z) H_A \rangle$. Then, due to this scattering process and the resulting shape deformation, the entanglement entropy of one sub-region would change as
\begin{gather}
\delta S_A = \int dx \sqrt{g} \  \nabla^\mu \zeta^\nu (x)  \langle T_{\mu \nu} (x) H_A \rangle + \mathcal{O} (\zeta^2).
\end{gather}

This equation should be compared with the equation 
\begin{gather}
\mathcal{A}_4 (s,t, u) = \int_{U_\Lambda}^ \infty dU \sqrt{g} \ e^{-2 \phi} \big( \prod_{i=1}^4 \psi_i (U) \big) \mathcal{A}_{10} (\tilde{s}, \tilde{t}, \tilde{u}).
\end{gather}
 
 From these two equations, one can see how the ten-dimensional scattering amplitude and the wave functions that are being scattered are related to the average of the correlation of the energy momentum and modular Hamiltonian. In other word, we claim that in an elastic scattering process we have the relation $\nabla^\mu \zeta^\nu (x)  \langle T_{\mu \nu} (x) H_A \rangle \propto e^{-2 \phi} \big( \prod_{i=1}^4 \psi_i (U) \big) \mathcal{A}_{10} (\tilde{s}, \tilde{t}, \tilde{u}) $.   This is consistent with the result of \cite{Jiang:2024noe}, where they showed that the scattering distance can be identified as the entanglement wedge cross-section. There they  also found connections between the string worldsheet and the Ryu-Takayanagi surface, and also mutual information and the geometric BV master equation \cite{Jiang:2024noe}. So the string scattering amplitude can be written in terms of the integration of the wave-functionals.

 This connection can further be checked phenomenologically by applying magnetic field. In \cite{Sonnenschein:2019bca}, it has been proposed that the magnetic field can be adjusted such that the amplitude vanishes for any kinematic setup. Similar studies in \cite{Ghodrati:2015rta, Dudal:2016joz} have been done for the case entanglement entropy, where it has been shown that for any size of subregion, the magnetic field can be tuned to make the entanglement entropy to get vanished. This then manifest further this connection.

Equation \ref{eq:poles} shows that the change in entanglement entropy only picks up the residue of the simple poles of $\langle T(z) H_A \rangle$ as
\begin{gather}
\langle T(z) H_A \rangle = ... + \frac{Res}{z - \partial A} + ... \ .
\end{gather}

 For the case of mutual information, as has been found in \cite{Chen:2022cmr}, the relation can be extended and the linear response of mutual information, $\delta I_{A,B}$,  can be encoded in the ``susceptibility" of shape deformation as
\begin{gather}
\frac{\delta I_{A, B} }{\delta \xi(X^i) } = (i) \oint_{ |z| = \tilde{z} } dz \bigg \langle \Big ( T_{zz} (z, \bar{z}, X^i) + T_{z \bar{z}} (z, \bar{z}, X^i)   \Big ) \Delta \tilde{H}_{A,B} \bigg \rangle + h.c,
\end{gather}
which picks up the residue of simple poles of $\langle T_{zz} \Delta \tilde{H}_{A,B} \rangle$ and is independent of the radial coordinate. This is exactly similar to the case of string scattering amplitude where the structure of the residues of its poles were discussed in \cite{Bianchi:2021sug}. So for the interval $A \equiv \{ -\frac{l}{2}  \le x \le \frac{l}{2} \} $, the modular flow which induces the change $\partial^\mu T_{\mu \nu} $ along $H_A$ follow a specific trajectory and a contour integral which is related to the trajectory of the strings that are being created from the vacuum, i.e, $\partial S_A \propto \oint_{z \sim \partial A} \langle T(z) H_A \rangle$.  This, as mentioned in \cite{Chen:2022cmr}, takes the residue of the simple pole in $\langle T(z) H_A\rangle$ where the poles that are emerging in the integral of the modular Hamiltonian are related to the poles of the string scattering amplitude $\mathcal{A}_4$.

Similarly, the scattering amplitude, $\mathcal{A}_4$, can be written as the integral of the metric, and the shape deformation due to the scattering process, and also in terms of the correlation functions between the stress tensor and modular Hamiltonian, i.e, $\langle T_{\mu \nu} \Delta H \rangle $. In terms of the modular flow, it can be written as $ \langle T_{\mu \nu} \mathcal{O} (-is_1) \mathcal{O}(-i s_2) \rangle$. Therefore, the logarithmic branch cuts observed in the string scattering process and its trajectories could be described by the modular trajectories and the zeros in the correlations between stress tensor and modular Hamiltonian. As mentioned in \cite{Chen:2022cmr}, in the shape deformation, the main contributing modes are the ones that change the position of the cut-off tube, instead of deforming it. So at low energies, these modes are the one that are related to the binding energy.

Actually, in \cite{Gao:2021tzr}, the scattering amplitude in terms of the modular flow has been written as
\begin{gather}
\mathcal{A} \left ( \{ \phi_l^i, \phi_r^j \} \to \{ \chi^k \} \right )= 
\text{Tr} \Big( \rho^{1-is} \chi^1 . . . \chi^n \rho^{is} \phi_l^1 ( t_{l1}) . . . \phi_l^i ( t_{li}) \phi_r^1 ( t_{l1}) . . . \phi_r^j ( t_{rj}) \Big),
\end{gather}
where $\{ \chi^k \}$ is the set of bulk operators which act on a thermofield double state, $\rho$ is the probe and $\phi_l^1 ( t_{l1}) . . . \phi_l^i ( t_{li}) \phi_r^1 ( t_{l1}) . . . \phi_r^j ( t_{rj})$ is the boundary state which generates the particles. So through this link between the scattering amplitude and modular flow, the direct relations between mutual information and scattering amplitude could be derived. Therefore, the information that scattering amplitude could catch such as detecting the binding energies at low $s$ could be caught by modular flow, entanglement entropy and mutual information as well. However, as explained before, the correlation of string amplitude with mutual information is more direct and physical compared with the entanglement entropy.

For example,  the mutual information between two spheres  can be written in a logarithmic function which has branch cuts \cite{Chen:2022cmr} and this gives signals for reasons behind the logarithmic branch cut behavior observed in string scattering amplitudes in \cite{Bianchi:2021sug}. 

In \cite{Chen:2022cmr}, actually, the linear response of the mutual information between the two spherical regions due to the scattering process or any other perturbative effects like shape deformation, has been written as
\begin{gather}
\delta I_{A,B}=N_d \left ( \frac{\partial I_{A,B}}{\partial \ \text{ln } \rho } \right ) \int d \Omega_{d-2} \left( \frac{L^2 - R_A^2}{L^2+ R_A^2 - 2 L R_A \cos \alpha} \right )^{d-1} \zeta ( \Omega).
\end{gather} 

This results in a form of susceptibility that resembles the integral used to calculate the scattering process amplitude. So one could see that
\begin{gather}
\int_{u_{KK}}^\infty du \prod_{i=1}^4 \psi_i (u) \propto \delta S \propto \delta^{-(d-2)} \int_{S_{d-2}} d\Omega_{d-2} \  \xi(\Omega) \nonumber\\ \propto  \delta^{-(d-2)} \int_{S_{d-2}} d\Omega_{d-2} \sum_{\ell, m} \xi^{(\ell, m)} \mathcal{J} (\Omega)^{-1} \Phi_{A, B}^{(\ell,m)} (\Omega).
\end{gather}

So the modes of the scattered strings can be written in terms of the decomposition modes of the deformation vector. This is consistent with the fact that the entanglement has a non-local structure.

Also, for the case of two spheres,  we could write the approximate relation
\begin{gather}
N_d \left( \frac{\partial I_{A, B} }{\partial \ \text{ln} \rho} \right) \left ( \frac{L^2 -R_A^2}{L^2+R_A^2-2 L R_A \cos \alpha } \right )^{d-1} \simeq \sqrt{-g} e^{-2 \phi} A_{10} \lbrack \tilde{s}(r), \tilde{t}(r), \tilde{u} (r) \rbrack,
\end{gather}
connecting the change of mutual information and $10d$ scattering amplitude.

Another point which is related to the interconnections between modular Hamiltonian and string scattering is the additive behavior in $s_k$, as the response of mutual information can be written as \cite{Chen:2022cmr}
\begin{gather}
\delta I_{AB}= \int ds_k ``\delta I_{AB} (s_k)".
\end{gather}

We can claim that this additive behavior arises from the connections between mutual information, modular flow, and the string scattering process.

\subsection{Time-dependent mutual information}
Another interesting direction to examine the connections between the spread of quantum information in spacetimes and string scattering amplitudes is the recent development in defining the space-time generalization of mutual information (STMI) and characterizing its behaviors as in \cite{Glorioso:2024xan, Fullwood:2024nrr, Milekhin:2025ycm, Chu:2025sjv}. 
In \cite{Glorioso:2024xan}, the space-time mutual information $J(A:B)$ has been defined by optimizing the relative entropy over the choice of gates $V_A$ and $V_B$ as 
\begin{gather}
J_N (A:B) = \frac{1}{N} \sup_{V_A, V_B}  S\left ( \sigma^{(N)}_{W,1} \Big | \sigma^{(N)}_{W,0} \right ), \ \ \ \ \ \ \ J(A:B) = \lim_{N \to \infty} J_N(A:B).
\end{gather}

These gates would actually act as the ancilla-system coupling where $W$ plays the roles of an idler which carries the information encoded in $A$ to the future where these gates and $W$ are arbitrary. This notion is based on the hypothesis testing interpretation of relative entropy \cite{Hayashi:2001} as the relative entropy actually provides the fastest rate at which one can identify a wrong hypothesis before taking the measurement. 

One could  first consider a black box, which in our study here is the confining geometry. In the box there are either $N$ copies of quantum state $\rho$ or $N$ copies of quantum state $\sigma$. Then one can do measurements on this $N$-copied system and determine whether it is $\rho^{\otimes}$ or $\sigma^{\otimes}$. If one makes a hypothesis that the state is $\sigma^{\otimes}$ and carry out the measurement, then the probability of a measurement outcome could be found by assuming the state is $\sigma^{\otimes N}$.  If the initial hypothesis is correct, then the probability will reach $1$ in the large $N$, but if it is incorrect, and the state be $\rho$, the probability $P_N(\rho  | \sigma)$ would be smaller. This probability would have a lower bound as $P_N (\rho | \sigma) \ge e^{- N S(\rho | \sigma)}$, where $S(\rho | \sigma) = \text{tr} ( \rho \log \rho - \rho \log \sigma)$. For two spacelike separated regions $A$, $B$, the mutual information is a relative entropy $I(A:B) = S ( \rho_{AB} | \rho_A \otimes \rho_B)$.

For the case where $\sigma$ is not full rank, there would be states which never appear in $\sigma$ and in this case the relative entropy would \textbf{diverge}. One could imagine a qubit system with states $ \ket{0}$ and $\ket{1}$. Then, if $\sigma = \ket{0}\bra{0}$, the probability of seeing $\ket{1}$ is zero, so as $P_N=0$ then $S(\rho | \sigma) \to \infty$. This case corresponds to the branch cuts behavior of mutual information as we have found in the previous section, which is due to creation of bound states.

The expression for $J_N(A:B)$ can also be written as \cite{Glorioso:2024xan}
\begin{gather}
J_N(A:B) = \sup_V \frac{1}{N} \Big \lbrack I^{(N)} (B: W) + S \left( \rho_{B^N} \big | \rho^{\otimes N}_{B,0} \right )\Big \rbrack.
\end{gather}

From the above relation, the condition for the divergence of relative entropy is that $\rho_{B,0}$ should not be full rank and $\rho_{B^N}$ has a nonzero probability to be in the null space of $\rho^{\otimes N}_{B,0}$. As explained in \cite{Glorioso:2024xan}, if in a system there is a conserved charge and the original state $\rho_{B,0}$ is supported in a charge range $\lbrack q_1, q_2 \rbrack$, then if one can tune $V$ to change the charge of $B$ to go beyond this range, then $J(A:B)$ would diverge. This exactly corresponds to the creation of bound states in the confining geometries, as there are fluxes and conserved charges which can change the charge of the subsystem $B$. This corresponds to a certain set of ancilla which has a finite probability to show up and can cause this effect. Also, such divergences have a logarithmic nature which is similar to the low-energy logarithmic singularities and logarithmic branch points of \cite{Bianchi:2021sug, Bianchi:2023uby}, which replace the poles in the original flat space amplitude.

Note that the full amplitude has the relation \cite{Bianchi:2021sug}
\begin{gather}
\mathcal{A}_4 (s,t,u) = \frac{r_0^4}{R^4} \frac{1}{4-\Delta} \sum_{n=0}^\infty \frac{\mathcal{R}_n(\theta)}{n+1} \  {}_2F_1\left(1; \frac{\Delta}{2}-2; \frac{\Delta}{2}-1; \frac{\alpha'_0 s}{4(n+1)} \right),
\end{gather}
where for even $\Delta = 2(k+1) >2$, and for any integer $k$, the hypergeometric function can be written as a logarithmic function as
\begin{gather}
{}_2 F_1 (1; k; k+1; z) = - \frac{k \log(1-z) }{z^k} - \sum_{l=1}^{k-1} \frac{k}{(k-l) z^l}.
\end{gather}

On the other hand, for factorized initial state, as found in \cite{Glorioso:2024xan}, the relative entropy could be written as 
\begin{gather}
S(\rho_{BW} | \rho_{BW,0} ) = -S(\tilde{N} ( \rho_W)) + S(\rho_W) - \text{Tr} \rho_W \mathcal{N}^\dagger \log \mathcal{N}(\rho_{\text{in}}),
\end{gather}
where $\rho_W$ could also be written as 
\begin{gather}
\rho_W = C \exp \left \lbrack \tilde{\mathcal{N}}^\dagger \log \tilde{\mathcal{N}} (\rho_W) - \mathcal{N}^\dagger \log \mathcal{N}, (\rho_{\text{in}}) \right \rbrack,
\end{gather}
and $\mathcal{N}$ here are quantum channels.  So the radial direction in confining geometries where the integral is being taken over could be related to the quantum channels to explain the Logarithmic divergences. The time evolution then can be defined by a quantum channel $\mathcal{N}$ and also the holographic direction can be mapped to the time direction where its integral could be considered as a time interval.

For unitary channels $\mathcal{N} : A \to B$, the quantities are 
\begin{gather}
\rho_W = \frac{\rho^{-1}_{\text{in} } }{\text{Tr}  \rho^{-1}_{\text{in} } }, \ \ \ \ \ \ \ \ \  J_1 = \log \text{Tr} \rho^{-1}_{\text{in}}.
\end{gather}

For the case of single-qubit system, the divergent part of relative entropy can be found from  \cite{Glorioso:2024xan},
\begin{gather}
S ( \rho_{BW} | \rho_{BW,0}) \approx - \text{Tr} \mathcal{N}_{\text{dph}} (\rho_W) \log \mathcal{N}_{\text{dph}} (\rho_{\text{in}}),
\end{gather}
where $ \mathcal{N}_{\text{dph}} $ is a dephasing channel. Then, the space-time mutual information (STMI) has been found as
\begin{gather}
J_1(A:B) \approx -2 \log \epsilon.
\end{gather}

The STMI diverges as the initial state approaches an \textbf{eigenstate} of the evolution and is independent of the expansion coefficients $0 \le p \le1$. This result has been generalized to many-body localized systems as well.  This getting approached to the eigenstate of the evolution is dual to the bound state creation.

For instance one could take $A$ as a single qubit at the initial time $t=0$ and $B$ the same qubit at time $t$, and initial pure states on $A$ would be of the form $\rho_{\text{in}} = \ket{\chi} \bra{\chi}$, where
\begin{gather}\label{eq:initialState}
\ket{\chi} = \cos \alpha \ket{0} + \sin \alpha \ket{1}.
\end{gather}

As the STMI actually quantifies the information preserved by the system, during bound state creation, one would expect that it diverges, since one can get closer to the initial state \ref{eq:initialState}. This is an eigenstate of the conserved operator $\sigma^3$, which getting closer to it would make the STMI larger.

In addition, in \cite{Milekhin:2025ycm}, a family of entanglement measures for time-separated subsystems based on von Neumann entanglement entropy or Tsallis entropy have been proposed. For a one-dimensional free fermion chain which has the Hamiltonian
\begin{gather}
H= \sum_s \psi(s) \bar{\psi} (s+1) + h.c,
\end{gather}
their measure for Euclidean complex coordinates $z=s+ i \tau$ and for two intervals $\lbrack z_1, z_2 \rbrack \cup \lbrack z_3, z_4 \rbrack$ would lead to
\begin{gather}
\text{Tr} \rho^n = \left ( \frac{\epsilon^4}{ x \bar{x} (z_2 - z_1) (\bar{z}_2 - \bar{z}_1)(z_4 - z_3)(\bar{z}_4 - \bar{z}_3) } \right )^{- \frac{n-1}{12n}},
\end{gather}
where
\begin{gather}
x= \frac{(z_4 - z_1)(z_3 - z_2) }{(z_3 - z_1)(z_4 - z_2)},
\end{gather}
and the numerical results matches with the CFT prediction which is
\begin{gather}\label{eq:STMICFT}
I_{\nu N} = \frac{1}{3} \text{Re} \log \left ( 1- \frac{L^2}{t^2} \right).
\end{gather}

\begin{figure}[ht!]
 \centering
   \includegraphics[width=12.5cm] {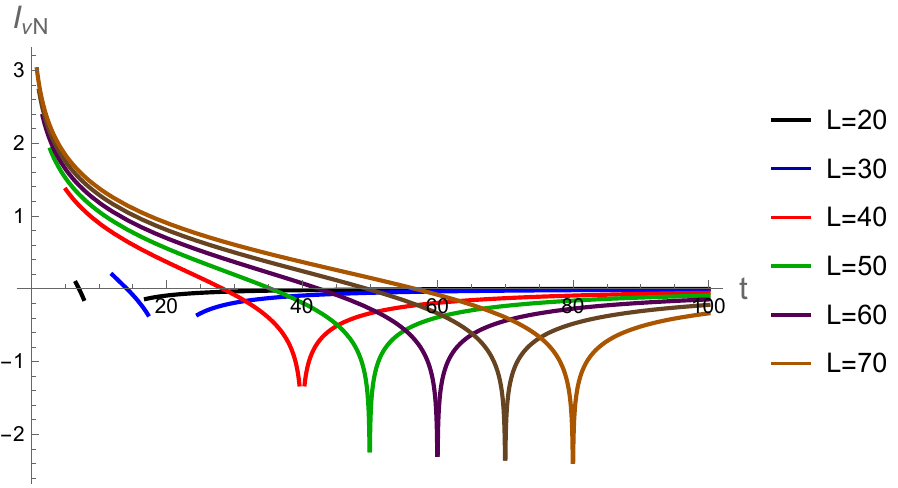} 
  \caption{The relation between time-dependent mutual information \ref{eq:STMICFT} for free fermions with various subsystem size $L$, which are separated by time $t$. }
 \label{fig:timeMIsystem}
\end{figure}

For different $L$, the relation between $I_{\nu N}$ and time $t$ has been shown in figure \ref{fig:timeMIsystem}.

\begin{figure}[ht!]
 \centering
   \includegraphics[width=5cm] {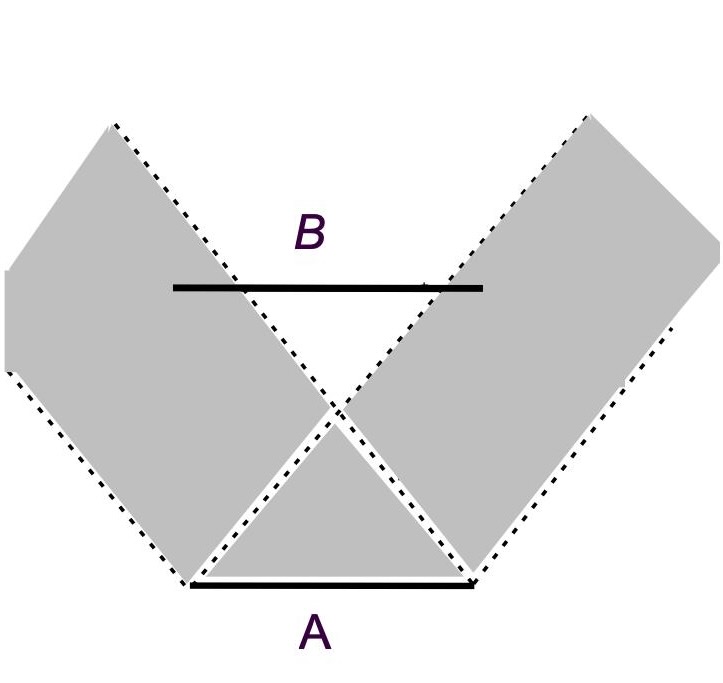} 
  \caption{The causal diamond between region $A$ and $B$ where the operators on $A$ can only have nonzero anti-commutators with the operators that are in the shaded region. So as time passes and the region $B$ gets outside of the shaded region, smaller parts of $B$ can be affected by $A$ and $s$ would decrease. }
\label{fig:diamondMutuals}
\end{figure}

It could be seen that increasing $L$ would increase the time that the branch-cut happens and it can magnify its peak, but the two branch become sharper itself. Note that the Mandelstam variable $s$ for a 2particle $\to$ 2particle process have the relation $s=(p_1 + p_2)^2=(p'_1+p'_2)^2$ where $p^2_1=m_1^2, \  p_2^2=m_2^2, \  {p'}_1^2 = {m'_1}^2, \ {p'}^2_2={m'_2}^2$. For bigger time interval $t$ between the two subsystems, the transformed energy and information and therefore $s$ would be smaller. This concept is shown in figure \ref{fig:diamondMutuals}.

Furthermore, in \cite{Fullwood:2024nrr}, using the pseudo-density matrix formalism, the timelike quantum mutual information has been defined where again the  ``branch-cut behaviors" could be noticed in their phase space, specifically in their $3d$ plots of mutual information between a single qubit at multiple times versus the depolarizing parameter $\eta$,  and the parameter $p$, $0<p<1$, which is a defining parameter of the initial state. 

So these analysis and the notion of time-like mutual information can shed more light on the connections between the quantum information behaviors and its spreading and the string scattering amplitude.

\section{Chaos, mutual information and string scattering amplitude}\label{sec:sec5}

In the studies of the chaotic behavior of highly excited (open) strings state (HES), decaying into two or three tachyons, \cite{Bianchi:2023uby, Firrotta:2023wem, Bianchi:2022mhs, Das:2023cdn, Savic:2024ock}, it has been shown that the peaks of the angular dependence of the amplitude exhibit characteristics comparable to the \textbf{$\beta$-ensemble} in random matrix theory, thereby confirming their chaotic nature. This chaos is due to the random superposition of various spherical harmonics in the partial wave expansion of the amplitude \cite{Bianchi:2023uby}. One would think that the signature of this chaos would be present in mixed quantum measures such as mutual information as well. In fact, the pattern of the successive peaks in the mutual information in the background of confining geometries which can be seen from figures \ref{fig:wedgewittenQCD}, \ref{fig:wedgeKT} and \ref{fig:SakaiDc}, would also point to this chaotic behaviors in the same phases where the damping of the peaks follows the same $\beta$-distribution.

\begin{figure}[ht!]
 \centering
   \includegraphics[width=8.3cm] {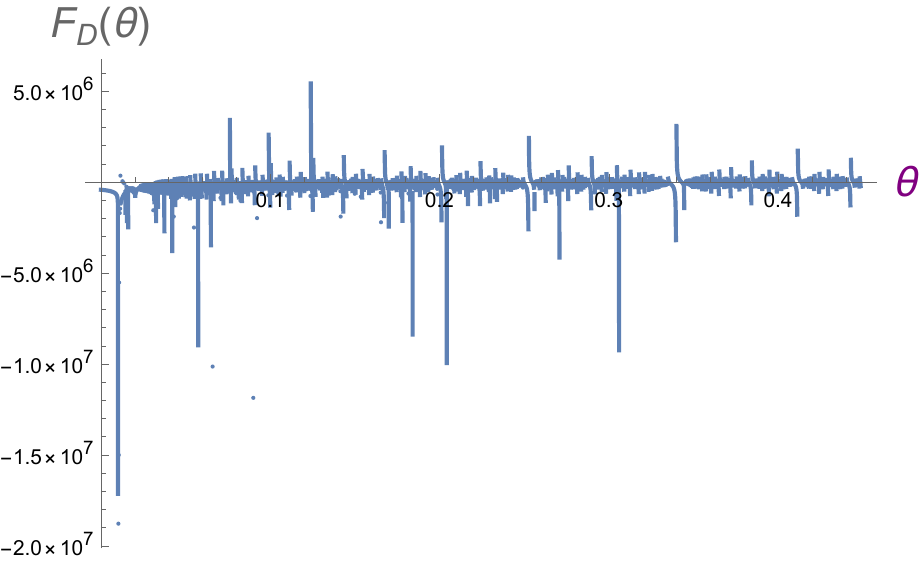} 
  \caption{The erratic behavior of the dressing factor $F_{\mathcal{D}} (\theta)$ vs scattering angle $\theta$, for $N=100$, $J=18$, \cite{Bianchi:2023uby}. }
 \label{fig:erraticFD}
\end{figure}

The observed chaotic behavior in the scattering amplitude of highly excited strings, linked to the string/black hole correspondence, offers valuable insights into the inherent chaos characteristic of black holes. In \cite{Bianchi:2023uby}, the authors demonstrate that multiplying the dressing factor of highly excited strings by the Veneziano amplitude results in a quantity that can interplay between them and create the transition from chaotic to regular spacing, as illustrated in their Figure 7 of  \cite{Bianchi:2023uby}. This could be written as 
\begin{gather}
\mathcal{A}_{HES} (s,t) = \mathcal{A}_{\text{Ven}} (s,t) \mathcal{D}_{HES} (s,\theta),
\end{gather}
where at the chaotic phase,  for the solution of $F_{\mathcal{D}} (\theta) \equiv \frac{d \log \mathcal{D}_{HES}  }{d\theta}=0$, the ratios of consecutive spacing would follow the $\beta$-ensemble distribution. Also, at the edge of chaos,  the peaks of the dressing factor $\mathcal{D}_{HES}$ is a complicated product of polynomials where the peaks are spaced erratically. An example, in the Regge limit, is shown in figure \ref{fig:erraticFD}.

One would expect that various quantum information measures would show the same structure and have a transition between regular and chaotic structures. For instance, in recent works such as \cite{Huh:2024lcm}, the Krylov complexity has been used to track the transition between the integrable and chaotic regimes which could also catch the eigenvalue statistics in the mixed phase systems. Therefore, it could be expected that the Krylov complexity can be connected to high energy string scattering amplitude as well.

In our work, by varying the critical distance between two mixed states in a confining background, we observed a detectable transition from regular to chaotic behavior in the response of mutual information, as illustrated for instance in Figure \ref{fig:AdSSoliton}. This suggests a correlation between the scattering amplitude and mutual information, with both exhibiting similar behavior at the transition point. To solidify this link, one could compare patterns in the peaks of mutual information with the results of random matrix theory (RMT), potentially through an angular functional between the two mixed points, to further strengthen the connection.

\begin{figure}[ht!]
 \centering
   \includegraphics[width=8cm] {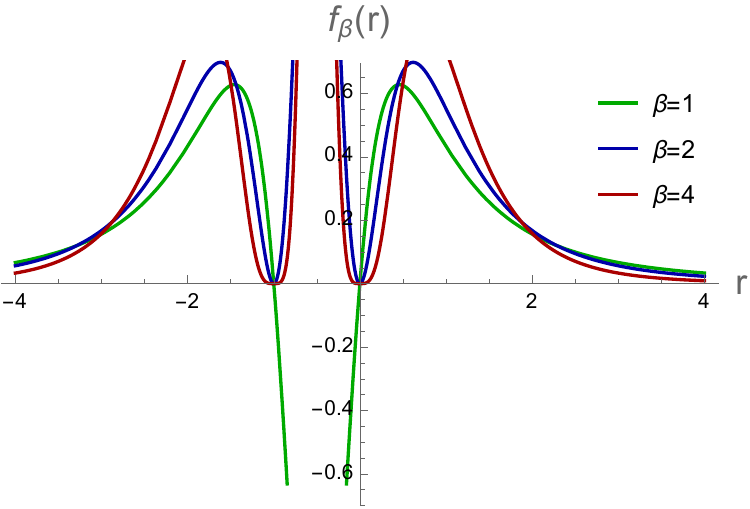} 
      \includegraphics[width=8cm] {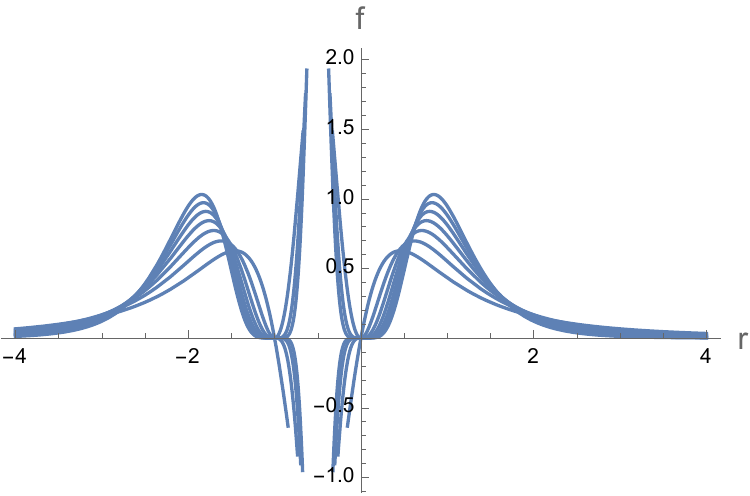} 
  \caption{The probability density functions for the ratio of eigenvalues in the $\beta$-ensemble which are being used as fitting models.}
 \label{fig:densityFun}
\end{figure}

In \cite{Bianchi:2023uby}, it was also demonstrated that for a $3 \times 3$ matrix with eigenvalues $\lambda_1 < \lambda_2 < \lambda_3$, the probability density function for the ratio $r= (\lambda_3 - \lambda_2)/(\lambda_2 - \lambda_1)$ in the  $\beta$-ensemble is given by
\begin{gather}
f_\beta (r) = \frac{3^{\frac{3+3\beta}{2} }\Gamma(1+\frac{\beta}{2} )^2 }{2\pi \Gamma(1+\beta)} \frac{(r+r^2)^\beta }{(1+r+r^2)^{1+\frac{3}{2}\beta}},
\end{gather}
which resembles the probability distribution function of the normalized mutual information in Multiple-input multiple-output (MIMO) systems and the energy levels of a Coulomb fluid \cite{DBLP:ShangLi}. The plot of distribution is shown in figure \ref{fig:densityFun}. It can be seen that by increasing $\beta$, the maximum in the distribution function becomes bigger and the peaks get further away from each other. Since the case of $\beta \to \infty$ corresponds to a rigid structure or crystallization, one might think that bigger $\beta$ corresponds to smaller mutual information between two mixed subsystems. This is indeed consistent with the result of \cite{Hern_ndez_2019}, where it is shown that large $\beta$ leads to conditional probabilities close to $1/2$ and small mutual information, while small $\beta$ makes conditional probabilities to be around the borders $0$ or $1$, leading to larger mutual information. This is also consistent with our result here about the connections between string scattering amplitude and mutual information.

The source of the chaotic behaviors and partonic structures observed in the scattering amplitude and mutual information is large fluctuations in the spin of the intrinsically highly excited states, i.e, the complex superposition of many states with different spin $s$. Such superposition would affect the mutual information and the spreading of quantum information in the hard wall confining model as well.  So, as mentioned in \cite{Bianchi:2022mhs}, this huge degeneracy of string excitations which have the same level $N$ with generic spins could create interesting chaotic structures.
The precise impact of these spin fluctuations on quantum information measures, particularly mutual information and other mixed correlation metrics, needs further studies. 

\begin{figure}[ht!]
 \centering
   \includegraphics[width=4cm] {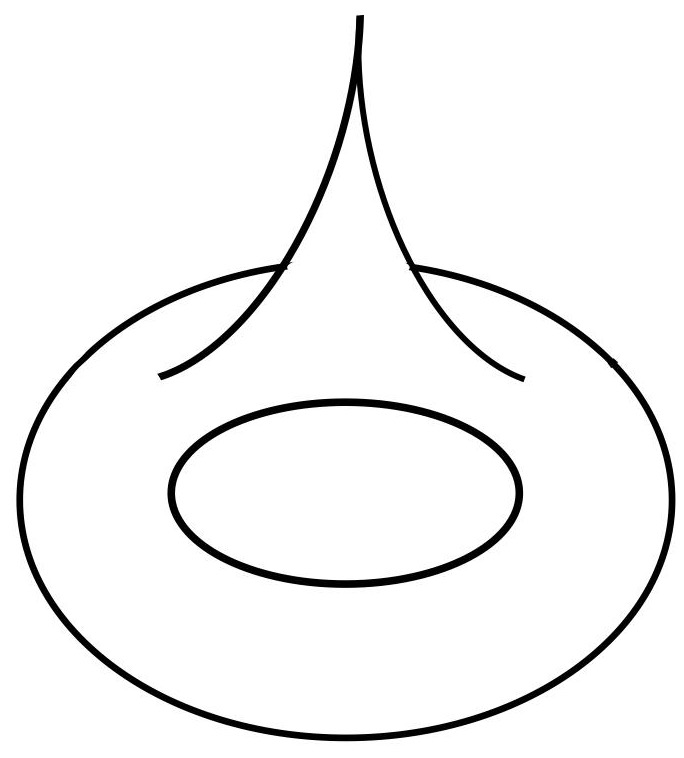} 
  \caption{The topology of a leaky torus embedded in Euclidean 3-dimension space \cite{GUTZWILLER1983341}. This geometry is smooth and orientable but not compact, with a constant negative curvature, and an infinite cusp. Also, it contains an open end where the phase shift of a particle which enters through it shows all the aspects of chaotic behavior, with great difficulty in its effective numerical computation. The chaos in this geometry could be compared with the chaos structures in confining backgrounds.}
 \label{fig:leakyBox}
\end{figure}

As discussed in \cite{Bianchi:2022mhs}, the successive peaks of the chaotic scattering amplitude are expected to follow a log-normal distribution. This log-normal behavior is also observable in the pattern of peaks in the mutual information as a function of the critical distance. Notably, the same metric can be applied to both the decay process of a highly excited string state into two tachyons and to quantum scattering on a ``leaky torus''. The topology of the leaky torus is illustrated in Figure \ref{fig:leakyBox}.

The geometry of the leaky torus $H$ can be written as \cite{GUTZWILLER1983341}
\begin{gather}
ds^2= \frac{dx^2 + dy^2}{y^2},
\end{gather}
where the boundaries of H are
\begin{equation}
\begin{aligned}
 (i) \ \  x = -1,  \ \ 0 < y < \infty; \ \ \ 
(ii) \ \  (x- \frac{1}{2})^2 +y^2 = \frac{1}{4}, \ \ 0<x<1; \nonumber\\ 
(iii) \ \  x =1, 0 < y < \infty ; \ \ \ 
(iv) \ \  (x+ \frac{1}{2})^2 + y^2  = ( \frac{1}{2})^2.
\end{aligned}
\end{equation}

So, the pattern of the scatterings of the wave-function on the leaky torus and the erratic behavior of its phase shift $\Phi$ as the function of momentum $k$, i.e, $\Phi(k) = \frac{\text{Im} \lbrack  \zeta (1+2 i k)  \rbrack }{ \text{Re} \lbrack \zeta(1+ 2 i k)   \rbrack } $ could be related to the spreading of information in a confining geometry with a hard wall. The spacings for successive maxima for both the function $\Phi (k)$ and $D_c (u_{KK})$ would be on the zeros and on the critical line $\text{Re} \lbrack s \rbrack = \frac{1}{2}$. The free wave scatters from $y = \infty $ and the phase shift between the incoming and outgoing waves is being measured at $y=y_0 >0$.

 \begin{figure}[ht!]
 \centering
   \includegraphics[width=8cm] {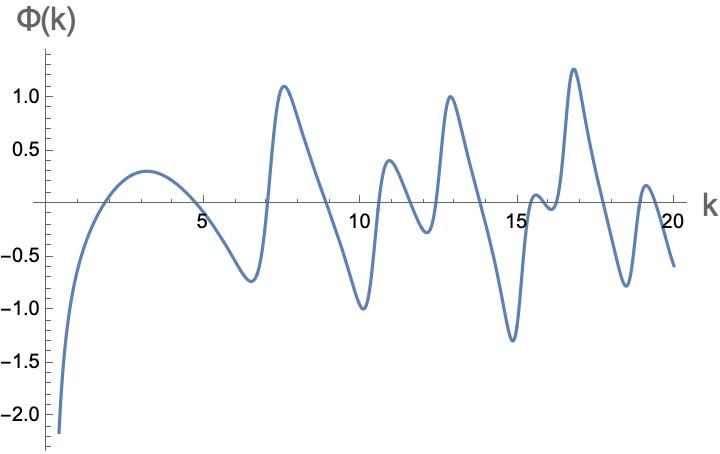} 
      \includegraphics[width=8cm] {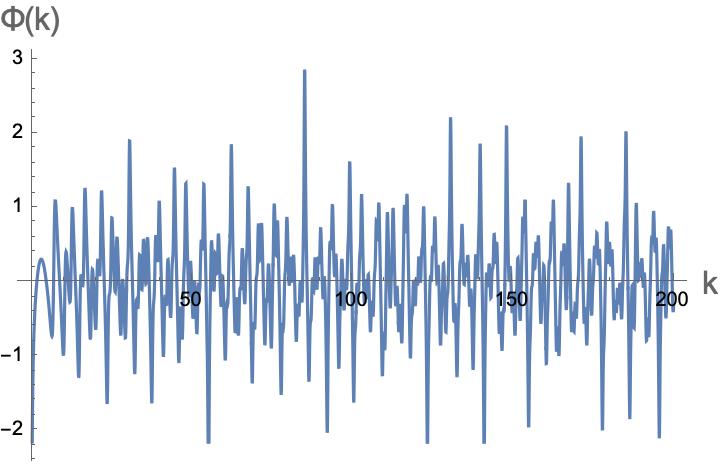} 
  \caption{The chaotic behavior of the phase shift as the function of the momentum of incoming wave on a leaky torus is shown here which can be a model of the spread of information in a confining geometry.}
 \label{fig:Phik}
\end{figure}

Then, the problems of quantum state spreading on a leaky torus, the decay of highly excited strings (HES) into two tachyons, and the mutual information spreading in the presence of a hard wall of a confining system would all become related and comparable to each other, each showing chaos in their behaviors, as the peaks of the parameters would all follow zeros of the Riemann zeta function \cite{GUTZWILLER1983341,Bianchi:2022mhs}.

\section{Mutual Information and Compton scattering}\label{sec:sec6}

To understand better the connections between the structures of entanglement entropy or mutual information of mixed systems on one side, and string scattering amplitude from the other side, it would be interesting to consider their changes during various other scattering processes, the elastic (such as Rayleigh \cite{Vinogradov:21} and Mie \cite{Schotland_2016}) scattering, or non-elastic (such as Brillouin \cite{yang2024eebrillouin}, Raman and Compton) scattering. 

There are several results from some other works about changing entanlement entropy during various scattering processes which match our results here. For instance, as shown in \cite{Schotland_2016}, the change of entanglement entropy versus various scaling length of the setup such as radius of spherical scatterer in Mie scattering is not monotonic but it behaves in an oscillatory way where totally it is increasing. This is consistent with the behavior of $D_c$ and mutual information that we observed in section \ref{sec:sec2}.

 \begin{figure}[ht!]
 \centering
   \includegraphics[width=7cm] {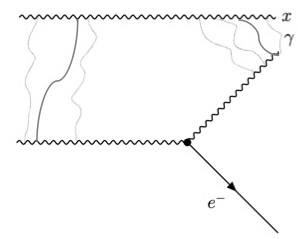} 
  \caption{The scattering process where a photon $\gamma$ scatters off an electron $e^-$ studied in \cite{Shivashankara:2023koj}. In the above, $x$ is a witness particle (photon) not participating in the process but is still entangled with the second photon. One can imagine that the two wiggly bosons are the two end points of an open string, and then calculate the change in the mutual information in this process, as relation \ref{eq:EEinScatter22}.}
 \label{fig:wiggly}
 \end{figure}

One of the interesting form of inelastic scattering which can be related to string scattering is Compton scattering where the change of entanglement entropy and mutual information during the process could be compared with the results we got in the previous sections. In fact, in \cite{Shivashankara:2023koj}, the entanglement entropy of the Compton scattering process in the presence of a witness has been discussed. In their setup, one can imagine that the two entangled photons are the end-points of a string, where one of them is being participated in the Compton scattering and the other one is a witness. The process is shown in figure \ref{fig:wiggly}.

One can imagine that the two wiggly bosons are actually the two end points of an open string moving on the worldsheet. Then, using this picture and repeating the procedures that have been done for the case of entanglement entropy as in \cite{Shivashankara:2023koj}, in terms of the Stokes parameter, one can calculate the change in the mutual information during this process. In low energy Compton scattering, the final mutual information of the electron and the witness particle's polarization is non-zero. This is due to the scattering process where in spite of no direct interaction between the two particles, they become correlated, indicating a correlation between the string scattering and mutual information.  

First, the initial pure state can be written as $\ket{i}= | e^- , r \rangle ( \cos \eta | \gamma, \uparrow \rangle + e^{i\beta} \sin \eta | \gamma, \downarrow \rangle | x , \uparrow \rangle )$, where $\eta$ is the entanglement parameter for the two photons $\gamma$, \cite{Shivashankara:2023koj, Fan:2017mth} ,  and $x$ can be considered the two end points of a string. Therefore, the parameter $\eta$ is proportional to the string tension $T$. The mass of the particles on the other hand is related to the energy of the string, which can be compared with the position of the hard-wall in the confining geometry.

In terms of $\eta$, the mass of the electron $e$ and the initial energy of photon, one can find the initial entanglement entropy of $\gamma$ or $x$, the change in the entropy of electron $e$, and the change of the entanglement entropy between $x$ and $e$ and therefore the total change of mutual information during this process.  

For example the change in the entanglement entropy between the coming electron $e$ and the photon at one end $x$ has been calculated by the relation  \cite{Shivashankara:2023koj}
\begin{gather}
\Delta S_{EE, ex}= -\frac{1}{2} ( 1+ 0.67 \frac{\sigma_T}{V/T} \frac{\omega^2}{m^2} ) \log \left ( 1+ 0.67 \frac{\sigma_T}{V/T} \frac{\omega^2}{m^2} \right)\nonumber\\
\ \ \ \ \ \ \ \   \ \ \ \  \ \  -\frac{1}{2} ( 1- 0.67 \frac{\sigma_T}{V/T} \frac{\omega^2}{m^2} ) \log \left ( 1- 0.67 \frac{\sigma_T}{V/T} \frac{\omega^2}{m^2} \right ),
\label{eq:EEinScatter22}
\end{gather}
where $\sigma_T = \frac{8 \pi r_e^2}{3} $ is the Thomson scattering cross section, $\omega$ is the initial energy of the photon, $V=(2\pi)^3 \delta^3(0)$ is the un-normalized volume, $m$ is the rest mass of the electron, and $T$ is a parameter in the S-matrix as in $S=1+i T$.  One can write the above relation and the change in the entanglement entropy and mutual information in terms of the Stokes parameter $\langle \sigma_z \rangle = - \frac{\sigma_T}{V/T} \frac{3 \omega}{2m}$ as well. In the above process, by increasing the ratio between the initial energies of $\gamma$ and the electron, i.e,.  $ (\omega/m)$,   the jump in the mutual information during the scattering process would increase. The rest mass $m$ here is directly related to the position of the end-wall of the confining geometry.

  \begin{figure}[ht!]
 \centering
   \includegraphics[width=6.5cm] {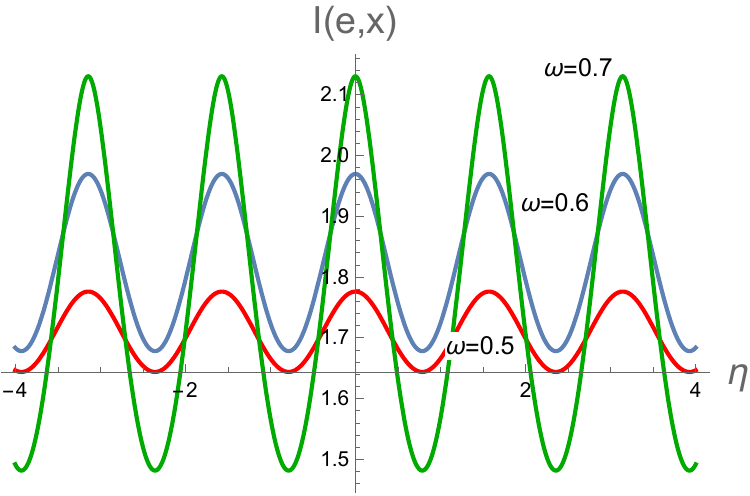} \ \ \ \ \ \ \ \ \ \
     \includegraphics[width=6.5cm] {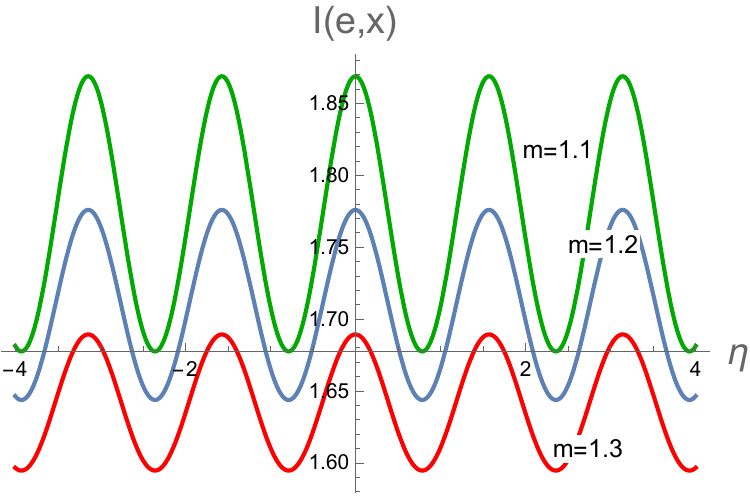} 
  \caption{The relation between the mutual information $I(e,x)$ between the two particles $\gamma$ and $x$,versus entanglement parameter $\eta$ for different mass and energy of photon. The $log$ are in based 2, so the mutual information are in bits.}
 \label{fig:SEELog2}
\end{figure}

For various energies $\omega$ and mass $m$ the behavior of $I(e,x)$ versus $\eta$ is shown in figure \ref{fig:SEELog2}. It can be seen that increasing $m$ and decreasing energy $\omega$ would suppress the mutual information as one would expect. making an analogy between hard wall and soft wall QCD, one could claim that if all parameters be the same, the soft wall leads to higher mutual information relative to the soft wall.

 One can also see an intrinsic periodic behavior in terms of the entanglement parameter here.  One can approximately simulate the process of strings hitting the end wall of the confining geometry with the scattering process of one photon hitting an electron and the other photon acting as a witness, which is actually the other end point of the string. Then, the change in the mutual information during this process can be found. Based on the connections between mutual information and string scattering amplitude, the periodic-like patterns we saw in certain phases, as in figures \ref{fig:wedgewittenQCD} and \ref{fig:wedgeKT}, could be expected.

Recently it has been shown that the fundamental string have a double-slit shape \cite{Hashimoto:2022ugt}, and the images of the strings via the Veneziano amplitude have been constructed. The information creating these images could also reveal the entanglement structures of such strings. As the scattering amplitude showed that strings have a double slit shape, the direct connection between the Veneziano amplitude and the mutual information of these two end-points of strings can be envisioned.

Also, in \cite{Hashimoto:2022ugt}, it has been shown that in the double-slit form of the strings, the slits are at the end points, which could be modeled by the system of figure \ref{fig:wedgeScatter}. This model can also help to reveal the highly complicated structure of a long string. In the confining backgrounds with a hard wall or soft wall, the chaotic behavior of strings leads to richer structures. So, the similarities between the fractal structure of the scattering amplitude and the quantum information measures could reveal further such complex structures.

\subsection{Two tachyons and one excited string amplitude}\label{subsec:sub6.1}

As another example to further explore the connections between string scattering amplitude and the spread of quantum information during the scattering process, one could consider the decay of a highly excited string into tachyons. Note that the probability of transitioning from initial states to the final state of the decay products is $P(f| i) \propto |\mathcal{A} ( i \to f)|^2$ and with these probabilities, one can construct the density matrix $\rho$ which describe the state of the system after scattering as $\rho= \sum_f P(f |i) \ket{f} \bra{f}$. The entropy then can be written as $S= - \sum_f P(f | i) \log P (f | i)$ which using it the mutual information between decay products of scattering processes such as tachyons can be computed. In a decay process of a highly excited string, the scattering amplitude determines the correlations between emitted particles where these correlations are encoded in the off-diagonal terms of the reduced matrix directly affecting the mutual information. In a chaotic system, the patterns of $P (f |i)$ are irregular and the mutual information can capture these irregularities which propagate into the correlations between the subsystems.

\begin{figure}[ht!]
 \centering
   \includegraphics[width=4.5cm] {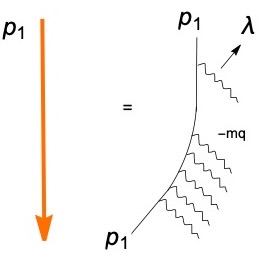}  \  \ \ \
     \includegraphics[width=4cm] {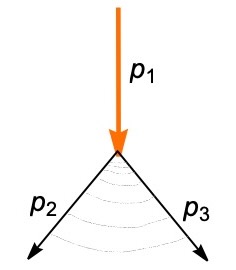} \ \ \ \
       \includegraphics[width=4.5cm] {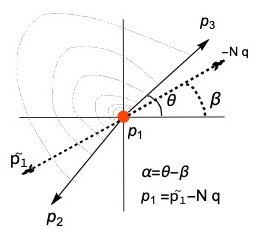}
  \caption{In the left, the DDF mechanism of exciting a heavy string with $n_m$ photons with momenta $-mq$ off of an initial tachyon with momentum $\tilde{p_1}$ is shown. In the middle, the structure of momentums are shown where the highly excited string has momentum $p_1$ decaying into tachyons with momenta $p_2$ and $p_3$. As time proceeds, the correlations among the decay products suppress. In the right part, the angles of different momentum vectors are shown as the amplitude depends on $\alpha= \theta-\beta$. The pattern of entanglement is also depicted. \cite{Gross:2021gsj} }
 \label{fig:ScatterTachyons}
\end{figure}

The excited string could be written as $\prod_{m=1}^\infty (\lambda \  .  \ A_{-m})^{n_m} \ket{0}$,  \cite{Gross:2021gsj}, where $N=\sum_{m=1}^\infty m n_m$, and the $m$th mode of a string is being excited $n_m$ times, i.e., by $n_m$ photons each with polarization of $\lambda$ and momentum $-m q$. Here $m$ is an integer and $q$ a null vector.  Then, the amplitude of the decay of this excited string into two tachyons can be written as  \cite{Gross:2021gsj}, 
\begin{gather}
\mathcal{A} \propto \prod_{m=1}^\infty (p_3 \ . \ \zeta P_m (p_3 \ . \ q))^{n_m}, \ \ \ p_3 \ . \ \zeta= - \sqrt{\frac{N}{2}} \sin \alpha, \ \ \ p_3 \ . \ q= - \cos^2 \frac{\alpha}{2}, \ \ \  P_m(a) = \frac{(1+m \ a)_{m-1} }{(m-1)!},
\end{gather}
where $\alpha$ is the relative angle between the tachyons' direction and the photons which create the string and $\lambda$ is the polarization of the DDF (Del Giudice, Di Vecchia, Fubini \cite{DelGiudice:1971yjh}) photons which are transverse to the momentum, i.e, $\lambda \ . \ q =0$. In addition, $(a)_{m} \equiv \frac{\Gamma(a+m) }{\Gamma(a)} = a (a+1) ... (a+m-1) $ is the Pochhammer symbol, unifying the rising and falling factorials. The setup is shown in figure \ref{fig:ScatterTachyons}. The correlations are depicted in gray lines. 

At early stages, due to the production of tachyons, the mutual information increases, and over time, as particles get apart and lose coherence, due to decoherence effects and causal separations the mutual information decreases.

As the string modes $n_m$ increase, the scattering amplitude and mutual information are also expected to increase. This is actually because, in a spacetime constructed from more entangled subsystems, the excited strings more easily interact with photons carrying the momenta $-m q$ or the null vacuum (vector) $q$.  Therefore, one would expect a power law relationship between the mutual information $I$ and the string modes $n_m$ as well.

  \begin{figure}[ht!]
 \centering
   \includegraphics[width=9cm] {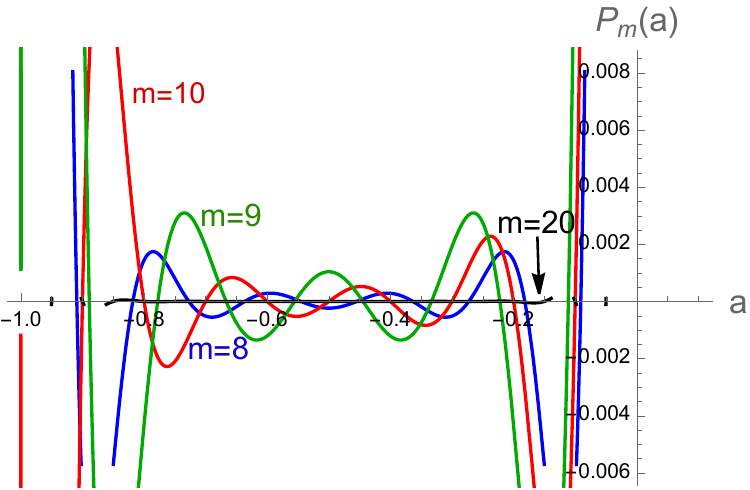}  
  \caption{The function $ P_m(a)$ of the Pochhammer symbol  $(a)_{m}$ for various $m$.}
 \label{fig:Pochhammer}
\end{figure}

Another interesting insight one can get from this formulation is on the source of the oscillatory behaviors in the mutual information. Based on the relation with scattering amplitude, this is rooted in the presence of the Pochhammer function of the Pochhammer symbol $(a)_m$ which its behavior is shown in figure \ref{fig:Pochhammer}. This function shows oscillatory behavior within the range of $-1 < a < 0$. As the parameter $m$ increases, the oscillations become more pronounced, even though the peaks of the amplitude suppresses. Additionally, for larger values of $m$, the function has more zeros and also the amplitude is influenced by $P_m(a)^{n_m}$. For large $m$, this amplitude can also be estimated by a product of several sine functions, resulting in a highly erratic behavior. As shown in \cite{Gross:2021gsj}, and by simplifying the Pochhammer function $\mathcal{A} \propto \prod_m (\sin (\pi m p_3 . q))^{n_m}$.

In the limit of large $m$ which corresponds to larger or more excited strings, a greater number of modes $n_m$ are stimulated, which in the confining case corresponds to bigger values of $u_{KK}$. Therefore,  one would expect that in larger values of $u_{KK}$, the behavior of mutual information or $D_c$ becomes more chaotic, as we have observed in figure \ref{fig:wedgewittenQCD}.  One could also expect the following approximate relation for the mutual information $ I \propto \prod_{q_b=1}^{q_l} P_{q_b}^{n_{q_b}}$, where $q_b$ is the number of quantum bits that are correlated in the mixed system and $q_l$ denotes the maximum number of qubits that can be incorporated into the subsystems. Therefore, mutual information is proportional to the Pochhammer symbol raised to the power of the number of qubits in the system.

By any change in the angle of the outgoing tachyons $\theta$, or $\alpha= \theta-\beta$, or the ingoing excited string $\beta$, the amplitude would show erratic behavior. In the confining case, changing $u_{KK}$ would also change significantly the angle of the hitting strings to the wall, or the angle of two subsystems relative to each other and how they ``see" each other \cite{Ghodrati:2022hbb, Ghodrati:2020vzm}. Also, both the string amplitude and the mutual information would be sensitive to the occupation levels of the propagating strings.

Another connections between entanglement entropy or mutual information and scattering amplitude can be derived from the results of \cite{aoude2024positivity} where the authors first showed the relation
\begin{gather}
\Delta \mathcal{E} \lbrack \Omega^{\text{prod}} \rbrack = 4 \left( \frac{1}{2 E_{k_1} 2 E_{k_2}} \frac{T}{V} \right) \text{Im} \mathcal{A}_{\alpha \beta}^{\alpha \beta} (k_1 k_2 \to k_1 k_2),
\end{gather} 
where $\Delta \mathcal{E}$ is the change of entanglement of two scalars which is being generated by scattering and can be related to mutual information.  Also, $| \Omega^{\text{prod}} \rangle$ is the product of unentangled initial state, and $\mathcal{E} \lbrack | \Omega \rangle \rbrack$ is the linearized entropy, given by $\mathcal{E} \lbrack | \Omega \rangle \rbrack = 1 - \text{Tr}_A \lbrack \tilde{\rho}^2 \rbrack$. In addition, $V \equiv (2\pi)^3 \delta_V^3 (0)$ and $T \equiv 2\pi \delta_T(0)$ are the divergences in the spacetime volume. In the confining geometries this factor is directly related to the position of the wall $u_{KK}$. The amplitude here is elastic in the internal quantum numbers. So, from the above relation, it was again expected to see the similar patterns for zeros and poles in both string scattering and mutual information, as we have noted in the first section. 

Then, using the above relation, it has been demonstrated that the positivity of the string scattering amplitude is directly equivalent to the positivity of entropy, expressed as
\begin{gather}
\text{Im} \mathcal{A} >0 \Longleftrightarrow  \mathcal{E} \lbrack | \Omega^{\text{prod}} \rangle \rbrack \ge0,
\end{gather}
 Therefore, one could imagine that further information from quantum information such as various inequalities in this field, specifically in different backgrounds such as confining ones, could point to further constraints and insights on scattering amplitudes.

\section{Kink scattering, chaos and fractal structure}\label{sec:sec7}

As another illustration of the connections between patterns of entanglement and mutual information, chaos and scattering, we consider the case of kink-antikink and kink-kink scattering processes for solitons and compactons which can model the exchange of information between subsystems and delocalization of localized information. Depending on the parameters of the system, there could be a phase with creation of new excitations such as breathers which is analogous to generation of entanglement, or a phase with separated kinks and antikinks corresponding to the decay phase. There is also a phase shift in the kink-antikink scattering which would mirror the way mutual information can retain memory of past interactions in a quantum system, while both have nonlinear dynamics with complex evolution in a non-additive way.

Also, in both mutual information and kink-antikink scattering, there is a threshold effect as the kinks and antikinks can only interact above a certain energy corresponding to a certain critical velocity.  Around this critical velocity the behavior becomes fractal \cite{Hahne:2023dic}.  This critical velocity corresponds to the specific $u_{KK}$ where the plots of $D_c$ in a confining background with an end-wall show the fractal behavior as well. This can be seen in our figure \ref{fig:wedgeKT}, where the fractal structures emerge around $u_{KK} \simeq 0.39$.

\begin{figure}[ht!]
 \centering
   \includegraphics[width=8cm] {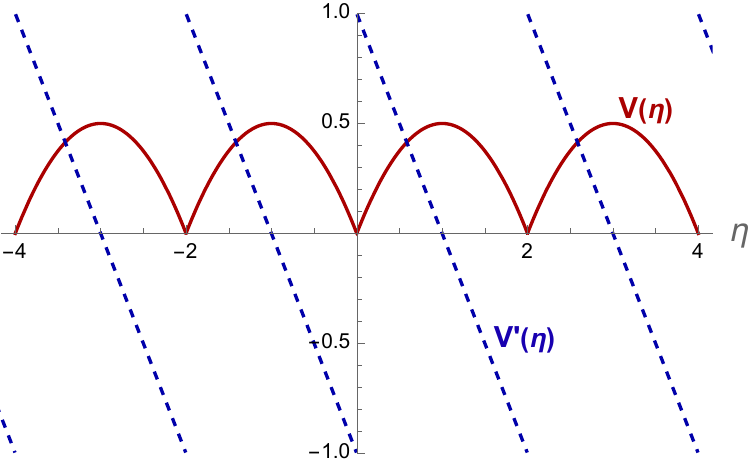} 
  \caption{The potential of a topological kink solution or a solitonic Skyrme model.}
 \label{fig: Skyrme}
\end{figure}

As an example of scattering in a topological kink solution, one first could consider a Heaviside-structured potential, similar to the one considered in \cite{Hahne:2023dic} as
\begin{gather}
V(\eta) = \sum_{n= - \infty}^\infty \left ( | \eta - 4n | - \frac{1}{2} (\eta -4n )^2 \right ) H_n ( \eta), \nonumber\\
H_n ( \eta) := \theta (\eta - 4n +2) \theta ( \eta - 4n -2),
\end{gather}
where its behavior and its derivative are shown in figure \ref{fig: Skyrme}. The confining solitonic Skyrme model can be considered as an example of having such potential. 
Here, $\eta$ is a real scalar field in $1+1$ dimension with the action
\begin{gather}
S= \int dt dx \left \lbrack \frac{1}{2} ( \partial_t \eta )^2 - \frac{1}{2} (\partial_x \eta)^2 - V (\eta) \right \rbrack,
\end{gather}
and with the BPS equation of $\frac{d\eta}{dx} = \pm \sqrt{2V(\eta) } $.

Note that this V-shape potential is a suitable for our analogy with QCD backgrounds, since this potential is also confining and can make particles localized at certain regions. If the height of this potential increases, because of its specific shape the subsystems become more spatially separated and therefore the mutual information would decrease. Also, this V-type potential would cause more decoherence and thermal fluctuations which also makes the mutual information to decrease.

Note that the most radiative case is shown to be the most chaotic. In the chaotic dynamics, the correlations would spread and get scrambled across the system and increase the mutual information. So as the energy levels $\delta_n = E_{n+1}-E_n$ or its ratios of successive spacing, and also the successive peaks of scattering amplitude $\mathcal{A}(\alpha)$ as a function of the continuous kinematical variable of the scattering problem $\alpha$ can act as diagnosis of chaos, with the analogy, the successive peaks of mutual information should play the same role as well. However, certain behavior of the system depends on the shape of the potential, for instance specific model of a holographic QCD model.

Since certain phenomenological and holographic QCD models, such as $V_{QCD}$, (see \cite{Gubser:2008yx,Ghodrati:2018hss}), have shapes similar to this V-shape potentials in certain regions, and as shown in \cite{Hahne:2023dic}, using collective coordinate method, one can similarly model mutual information in those systems using an internal mode as 
\begin{gather}
I_s= I_k \left(b(x-a)+ \frac{\pi}{2} \right) + c \chi \left(b(x-a)\right),
\end{gather}
where similar to the kink model, the internal mode can be written as
\begin{gather}
\chi(x) = 
\begin{cases}
\sin(2x) + \frac{1}{2} \sin(4x)& \text{if}   - \frac{\pi}{2} < x < \frac{\pi}{2} \\
0& \text{otherwise}.
\end{cases}
\end{gather}

\begin{figure}[ht!]
 \centering
   \includegraphics[width=6.5cm] {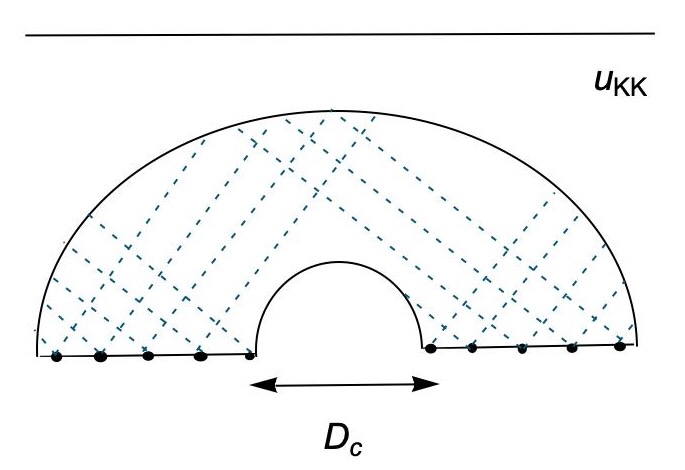}  
  \caption{Modeling the spread of mutual information between two mixed subsystems by kink-antikink scattering by analogy with causality cones. This could point to the use of internal modes to describe the chaotic regions in our plots at specific parameters.}
 \label{fig:diamond}
\end{figure}

Here $a$ is the position of the center of momentum of the kink, $b$ is a new coordinate to model the Lorentz contraction, and $c$ is the amplitude of the internal mode which can act as an additional coordinate. So, one  expects that adding this new internal coordinate, such as Derrick mode, can help to model the spread of mutual information in various confining models with potentials which has similarities with the V-shape model here. It has been shown that, in models with non-analytical potentials, the radiation is dominated by compact oscillons. Therefore, we could claim that in our setup, the radiation and the spread of mutual information in the confining backgrounds which have non-trivial potential would be dominated by the oscillons as well. In fact, taking into account the causality and light-cone structures for the spread of information from each qubits in each of the subsystems, a structure similar to figure \ref{fig:diamond} could be imagined which is similar to the shockwaves of  \cite{Hahne:2023dic}.

In \cite{Hahne:2023dic}, it has been observed from their figures 4 and 5 that the regions with most emission of radiation match the regions of chaotic behavior. These two are separated regions, one at smaller velocity $v$ and with a smaller peak, and the other bigger one at bigger $v$ and bigger peak. These two plots match with the behavior of mutual information or $D_c$, specifically in the case of Sakai-Sugimoto model.

Generally, in the interactions between the compact topological defects, which can approximately model the spread of information in confining geometries with an end-wall, two main classification classes for the scattering could exist. These are the kink-antikink pairs which oscillate and radiate, and the kink-antikink pairs which themselves emerge from the collisions where a shockwave is being formed, and this shockwave would decay into a ``cascade of compact oscillons" \cite{Hahne:2023dic}. These are in fact the zeros that the authors of \cite{Bianchi:2023uby,Bianchi:2022mhs} observed, demonstrating chaos in the scattering process.

 So if the position of the wall, temperature, energy, or other parameters of the QCD system get perturbed, one could expect the oscillons would be emitted in a similar chaotic pattern. This would also be similar to the scattering of kink and anti-kink or the scattering of compact oscillons in the signum-Gordon model. In our setup \ref{fig:wedgeScatter}, the shockwave borders would also act as energy or information reservoir.

  \begin{figure}[ht!]
 \centering
   \includegraphics[width=5.5cm] {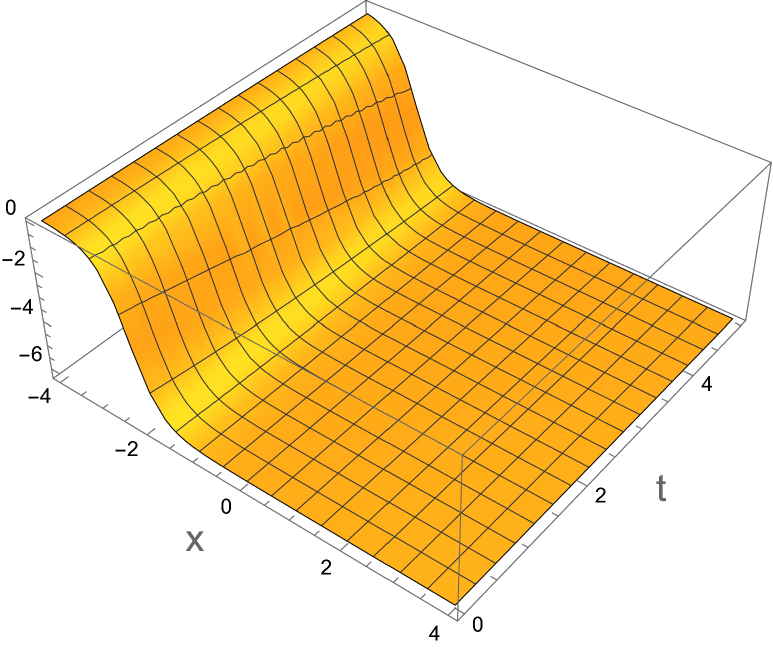} \ \ \ \ \ \ \ \ \ \ \ \ \ \  \ \ \ \ 
      \includegraphics[width=5.5cm] {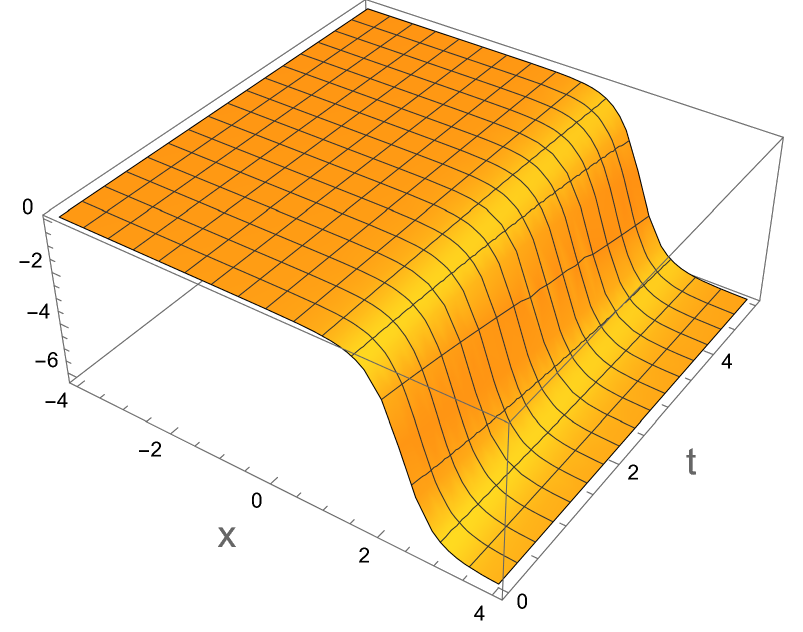} 
  \caption{The kink solution, given by the equation $u(x,t)=-4 \arctan (\exp(\lambda x + \frac{t}{\lambda} + \mu) )$ with $\lambda>0$, is shown on the left, while the anti-kink solution, with $\lambda<0$ is shown on the right part. The interaction and scattering of kink-antikink pairs can be compared to the spread of mutual information in confining models, where in both cases, critical phase transitions occur, and around it the fractal structures emerge. The appearance of periodic behaviors within these fractal-like structures has been observed in both cases.}
 \label{fig:etaV}
\end{figure}

The periodic relations between the kink-antikink parameters $\lambda_1$ and $\lambda_2$ and their velocities could be caught from the relation 
\begin{gather}
u(x,t) = 4 \arctan \left \lbrack  \left (\frac{\lambda_1 + \lambda_2}{\lambda_1 - \lambda_2} \right) \tan \left (\frac{v_1 - v_2}{4} \right )    \right \rbrack,
\end{gather}
where from the tangent function, the periodic structure could be detected.  Periodic structures in the zeros of string scattering amplitude appear when the highly excited string states involved are non-generic, as shown in figure \ref{fig:etaV}, such as the states in the first Regge trajectory.

  \begin{figure}[ht!]
 \centering
   \includegraphics[width=7cm] {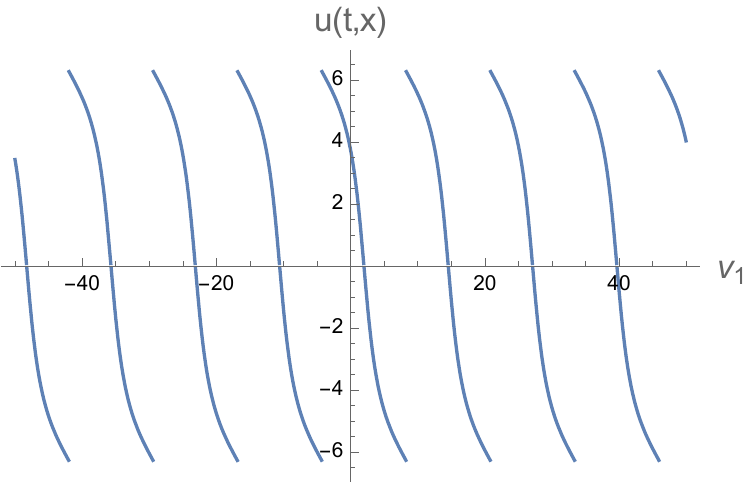}  \ \ \ \ \ \ \ \ \ \ \ \ \
      \includegraphics[width=6cm] {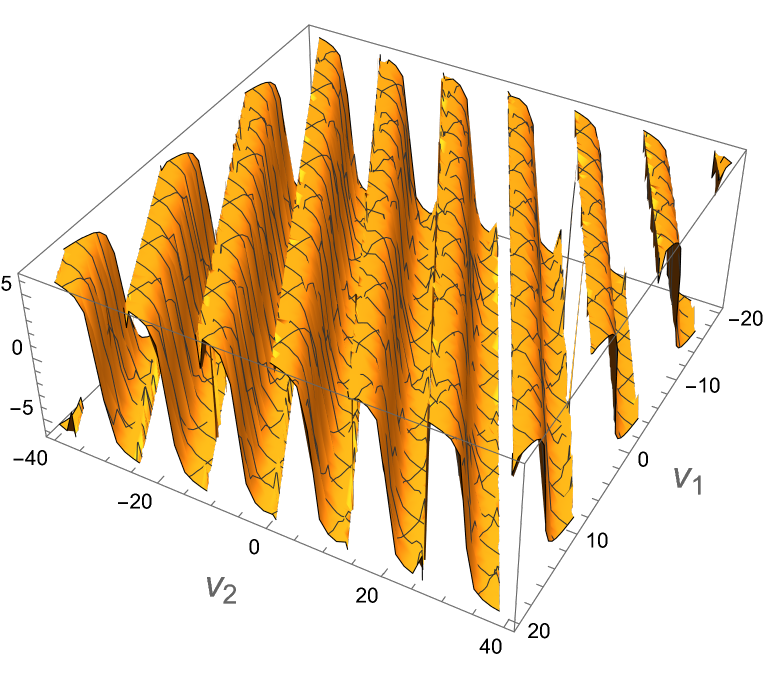} 
  \caption{The periodic structure in the behaviors of kink and anti-kink velocities is shown,  which can help to model the periodic structures in the behavior of mutual information or $D_c$ and also string scattering amplitude.}
 \label{fig:etaV}
\end{figure}

An interesting point is that higher scattering velocities result in smaller shockwaves and lower radiation intensities, while slower velocities produce the opposite effect. The oscillating behavior is only observable for velocities below a critical threshold, $v < v_c$. This explains the noise in the numerical studies of mutual information when the graviton considered to be massive and the scattering velocities get suppressed.  The increased noise also appears in the plots of entanglement entropy, complexity and complexity of purification, \cite{Ghodrati:2019hnn}.  The critical velocity, $v_c$, depends on various parameters of the model, such the graviton mass $m$,  particle charge $q$, background dimensionality $d$ and the position of the wall $u_{KK}$. This critical velocity is the initial velocity of kinks where a single bounce scattering would change to multi-bounce or the annihilation. 

Additionally, as mentioned in \cite{Hahne:2023dic}, close to the critical velocity $v_c$,  the escape and no-escape cases alternate in a fractal manner, corresponding to the fractal structures around $D_c$, where the mutual information between the subsystems drop to zero. The ``collective coordinate model" could be applied to describe all these collision processes and also the structures of quantum information and entanglement.

\section{Conclusion}\label{sec:sec8}

In this work, first, we showed in holographic confining backgrounds, and at low Mandelstam variable $s$, there are connections between the logarithmic branch cut singularities of the closed string scattering amplitude, with the behaviors of the peaks of mutual information or critical distance $D_c$. These peaks are the result of the dependence of the string tension on the holographic radial coordinate and are a sign of the creation of bound states at low energies. The statistics of these peaks demonstrate chaos in both cases. One can imagine that the two mixed subregions spread quantum information across the spacetime in ``modular time".  As the modular wavelets interfere with each other, and after hitting the wall, they create the chaotic structures. The power-law decay of both quantities match at higher energies. They also match based on the radial coordinate, as the real part of both follow the $s^{-1}$ curve. These observations further reinforces the connections between the patterns of entanglement entropy and string scattering amplitude.

To explain specific features of our results and elucidate the behavior of entanglement patterns, we implemented the modular flow and modular Hamiltonian as the foundation for introducing the ``modular time". Then, we drew connections between the scattering amplitude, binding energy and the spread of mutual information in these setups. Specially, we reviewed the new developments in defining space-time generalization of mutual information where again the same logarithmic branch-cut behaviors both in the analytical and numerical studied have been found and we connected these new concepts with the branch-cut behaviors found in the string scattering amplitude. These results could be helpful in using the spreading of quantum information in analyzing the behaviors of high energy strings.

Additionally, we discussed how mutual information in a setup with an end-wall, similar to the tripartite information can detect chaos. The leaky torus model has also been used to simulate the spread of quantum information and the emergence of chaos in different confining backgrounds which has an end-wall in their IR affecting the chaotic structures.

In addition, the exchange of entanglement entropy and mutual information during Compton scattering of two photons and also the decay of a highly excited string into two tachyons were explored and the source of the periodic structures were explained. Finally, the scattering of kink-antikinks,  the emergence of chaos in this scenario and the links with the behaviors noted in previous sections were discussed.

\section{Data Availability Statement}
Data Availability Statement: No Data associated in the manuscript.

\section*{acknowledgments}
I would like to thank  Dorin Weissman, Jacob Sonnenschein and Mao Tian Tan for useful discussions. This work has been supported by an appointment to the JRG Program at the APCTP through the Science and Technology Promotion Fund and Lottery Fund of the Korean Government. It has also been supported by the Korean Local Governments - Gyeongsangbuk-do Province and Pohang City - and by the National Research Foundation of Korea (NRF) funded by the Korean government (MSIT) (grant number 2021R1A2C1010834).

 \medskip

\bibliography{Amp.bib}
\bibliographystyle{JHEP}
\end{document}